\documentclass[aps,pre,reprint,groupedaddress]{revtex4-2}

\usepackage{amsmath}
\usepackage{amssymb}
\usepackage{bm}
\usepackage{hyperref}

\usepackage{graphicx}

\renewcommand{\vec}{\bm}

\begin{document}


\title{Laminar chaos in systems with random and chaotically time-varying delay}


\author{David Müller-Bender}
\email[]{david.mueller-bender@mailbox.org}
\affiliation{Institute of Physics, Chemnitz University of Technology, 09107 Chemnitz, Germany}
\author{Rahil N. Valani}
\affiliation{Rudolf Peierls Centre for Theoretical Physics, Parks Road, University of Oxford, OX1 3PU, United Kingdom}



\date{\today}

\begin{abstract}
A type of chaos called laminar chaos was found in singularly perturbed dynamical systems with periodically [Phys. Rev. Lett. 120, 084102 (2018)] and quasiperiodically [Phys. Rev. E 107, 014205 (2023)] time-varying delay.
Compared to high-dimensional turbulent chaos that is typically found in such systems with large constant delay, laminar chaos is a very low-dimensional phenomenon.
It is characterized by a time series with nearly constant laminar phases that are interrupted by irregular bursts, where the intensity level of the laminar phases varies chaotically from phase to phase.
In this paper, we demonstrate that laminar chaos, and its generalizations, can also be observed in systems with random and chaotically time-varying delay.
Moreover, while for periodic and quasiperiodic delays the appearance of (generalized) laminar chaos and turbulent chaos depends in a fractal manner on the delay parameters, it turns out that short-time correlated random and chaotic delays lead to (generalized) laminar chaos in almost the whole delay parameter space, where the properties of circle maps with quenched disorder play a crucial role.
It follows that introducing such a delay variation typically leads to a drastic reduction of the dimension of the chaotic attractor of the considered systems.
We investigate the dynamical properties and generalize the known methods for detecting laminar chaos in experimental time series to random and chaotically time-varying delay.
\end{abstract}


\maketitle


\section{Introduction}
\label{sec:introduction}

Whenever a phenomenon involves a transport process over a finite distance with a finite velocity it can be characterized by a time delay and therefore may be modeled by a time-delay dynamical system.
Such systems have been studied over decades and appear naturally in many fields of science \cite{kuang_delay_1993,scholl_handbook_2007,erneux_applied_2009,lakshmanan_dynamics_2011,scholl_control_2016} and engineering \cite{erneux_applied_2009,stepan_retarded_1989,michiels_stability_2007}.
An overview of recent developments can be found in the theme issues introduced by \cite{just_delayed_2010,erneux_introduction_2017,otto_nonlinear_2019} and a nice review on chaos in time-delay systems can be found in \cite{wernecke_chaos_2019}.
Despite the extensive research done in this area, which is also backed up by a well developed mathematical theory \cite{hale_introduction_1993,diekmann_delay_1995,hale_dynamics_2002}, fundamental questions are still open.
One of them is the influence of a variable delay on the dynamics of time delay systems.
In general, the delay generating process and therefore the delay itself can depend both on time and on the state variables of the system.
Exemplary systems with state-dependent delay can be found in regenerative turning models \cite{insperger_state-dependent_2007}, embryonic cell cycle oscillator \cite{rombouts_delay_2018}, Langevin equations with state dependent delay arising in the analysis of collective systems \cite{ge_data-driven_2024}, electrodynamics \cite{driver_two-body_1963,lopez_electrodynamic_2020}, and quantum analogs \cite{lopez_unpredictable_2024}.

If the state-dependence is weak, the delay may be approximated by a purely time-varying delay.
Depending on the structure of the system, such delays can stabilize \cite{madruga_effect_2001,otto_stability_2011,otto_application_2013} and destabilize systems \cite{louisell_delay_2001,papachristodoulou_stability_2007}.
It increases the complexity of systems \cite{radons_complex_2009,lazarus_dynamics_2016}, which was suggested to improve the security of chaos communication \cite{kye_characteristics_2004,ghosh_synchronization_2007,kye_information_2012}.
 Different types of synchronization caused by a time-varying delay were found in \cite{kye_synchronization_2004,kye_synchronization_2004_2,ambika_anticipatory_2009,ghosh_generalized_2009,ghosh_projective-dual_2011,senthilkumar_delay_2007,khatun_synchronization_2022}.
Time-varying delays were considered in the context of delayed feedback control \cite{gjurchinovski_stabilization_2008,gjurchinovski_variable-delay_2010,jungling_experimental_2012,gjurchinovski_delayed_2013} and on amplitude death in oscillator networks \cite{gjurchinovski_amplitude_2014}.
On a more general level, a temporal delay variation influences mathematical properties such as analyticity of solutions \cite{mallet-paret_analyticity_2014}.

While systems with slow and fast time-varying delays can be treated via approximation methods that lead to time-independent delays \cite{michiels_stabilization_2005,otto_application_2013}, a general delay variation apart from these limits leads to complex structures that are not captured by constant delay systems.
In order to understand the influence of a general delay variation on time-delay systems, we follow a bottom up approach, where we began with a periodically time-varying delay.
We then stepwise increase the generality of the delay variation using the insights of the preceding step until we reach the most general case of a state-dependent delay, which is our long time goal.  
General systems with a single periodically time-varying delay were considered in \cite{otto_universal_2017,muller_dynamical_2017}.
It turned out that there are two classes of time-varying delays, which lead to fundamental differences in the dynamics of the involved system as highlighted by drastic differences in the scaling behavior of the Lyapunov spectrum between these classes.
The first class, so-called \emph{conservative delay}, are equivalent to constant delays and therefore share their dynamical characteristics, whereas the second class, so-called \emph{dissipative delays} exhibit fundamentally different characteristics.

Using this framework, a prior unknown type of chaos called \emph{laminar chaos} was found in \cite{muller_laminar_2018,muller-bender_resonant_2019}, which is only observed in systems with dissipative time-varying delay.
Its existence and robustness was experimentally verified first in \cite{hart_laminar_2019,muller-bender_laminar_2020}, where an optoelectronic system was considered.
Further experimental realizations were achieved via electronic systems \cite{jungling_laminar_2020,kulminskii_laminar_2020}.
The synchronization of laminar chaotic systems was analyzed in \cite{khatun_synchronization_2022}, and in \cite{kulminskiy_laminar_2024}, for the first time, laminar chaos was found in a constant delay system coupled to a laminar chaotic time-varying delay system.
With laminar chaos we found a very pronounced physical phenomenon, which enables us to analyze the influence of a variable delay in a very vivid way.
Therefore, as a next step we considered laminar chaos in systems with the more general quasiperiodically time-varying delays \cite{muller-bender_laminar_2023}.
Such delays are relevant, for example, in the analysis of quasiperiodic solutions of systems with state-dependent delay \cite{he_construction_2017,he_construction_2016}.
In this paper, we consider random delay variations, which are common in many systems \cite{verriest_stability_2009,krapivsky_stochastic_2011,gomez_stability_2016,qin_stability_2017,liu_stability_2019} and are a natural limit of quasiperiodic variations while sending the number of frequencies to infinity \cite{muller-bender_laminar_2023}.
The present analysis of the strongly related chaotic delay variations is an important step towards understanding the influence of a variable delay on chaotic systems since a chaotic variation of the state implies a chaotic variation of the state-dependent delay.

We consider systems with dynamical structure
\begin{equation}
	\label{eq:sys}
	\frac{1}{\Theta} \dot{z}(t) + z(t) = f(z(t-\tau(\vec{h}(t))))
\end{equation}
where the variable delay $\tau(\vec{h}(t))$ depends on the output $\vec{h}(t)$ of an independent delay generating process.
Systems with constant delay, $\tau(\vec{h}(t)) = \tau_0$, have been extensively studied, where the choice of nonlinearity $f$ of the feedback depends on the specific application.
The nonlinearity $f(z)=\mu\,z/(1+z^{10})$ characterizes the Mackey-Glass equation \cite{mackey_oscillation_1977} which is a blood production model.
A sinusoidal nonlinearity, $f(z)=\mu\,\sin(z)$, gives the Ikeda equation \cite{ikeda_multiple-valued_1979, ikeda_optical_1980}, which is a model for light dynamics in a ring cavity with a nonlinear optical medium and describes certain optoelectronic oscillators \cite{hart_laminar_2019,larger_complexity_2013,chembo_optoelectronic_2019}.
General properties can be derived using a simpler quadratic nonlinearity $f(z)=\mu\,z(1-z)$ \cite{adhikari_periodic_2008}.
Normal and anomalous diffusion were found in systems with the climbing-sine nonlinearity $f(z)=z + \mu\,\sin(2\pi\, z)$ \cite{albers_chaotic_2022,albers_antipersistent_2022} and the double-sine nonlinearity $f(z)=z - \mu [\sin(2\pi\, z)+ (1/2) \sin(4\pi\, z)]$ \cite{albers_weak_2024}, respectively.
The parameter $\Theta$ sets the overall timescale of the system.
If $\Theta$ is large such that the short time scale $(1/\Theta)$ of the system is much smaller than the delay, Eq.~\eqref{eq:sys} belongs to the class of singularly perturbed delay differential equations and large delay systems, which are widely studied for constant delay, where slowly oscillating periodic solutions \cite{ikeda_successive_1982,chow_singularly_1983,mallet-paret_global_1986,adhikari_periodic_2008}, high-dimensional chaotic dynamics \cite{farmer_chaotic_1982,ikeda_high-dimensional_1987}, chaos control \cite{mensour_chaos_1998}, spatio-temporal phenomena \cite{yanchuk_spatio-temporal_2017} and chaotic diffusion \cite{albers_antipersistent_2022} have been observed.  
Also results on systems with state-dependent delay are available \cite{mallet-paret_boundary_1992,mallet-paret_boundary_1996,mallet-paret_boundary_2003,kashchenko_local_2015,martinez-llinas_dynamical_2015}.
The potentially high-dimensional dynamics of systems with the structure of Eq.~\eqref{eq:sys} and similar singularly perturbed systems is interesting for applications such as chaos communication \cite{goedgebuer_optical_1998,vanwiggeren_optical_1998,udaltsov_communicating_2001,keuninckx_encryption_2017}, random number generation \cite{uchida_fast_2008,reidler_ultrahigh-speed_2009,kanter_optical_2010}, and reservoir computing \cite{appeltant_information_2011,larger_high-speed_2017,hart_delayed_2019,stelzer_performance_2020}.

In this paper, we generalize the concept of laminar chaos to systems with random and chaotically time-varying delays.
The generic delay variations used for numerics are defined in Sec.~\ref{sec:delay_definition}.
After reviewing the theory on laminar chaos in Sec.~\ref{sec:review}, including the concept of conservative and dissipative delays for (quasi-)periodic delays, their generalizations are provided in Sec.~\ref{sec:lamchaos_rand}.
We first numerically analyze the dynamics for short-time correlated random and chaotically time-varying delays, where we present our main finding: Low-dimensional laminar chaos and its generalizations can be found in almost the whole delay parameter space spanned by mean and amplitude of the delay, whereas high-dimensional turbulent chaos is suppressed.
The results are then explained and verified by analytical considerations, which are exact in the large delay limit $\Theta\to\infty$, using the theory of circle maps with quenched disorder provided in \cite{muller-bender_suppression_2022}.
In particular, we find that the conditions for (generalized) laminar chaos presented in Sec.~\ref{sec:review} are still valid for random and chaotically time-varying delays because relevant physical quantities are unique for almost all realizations.
Our main finding is confirmed by the fact that these conditions are fulfilled if the minimum of the delay exceeds the correlation length of the delay variation leading to (generalized) laminar chaos above that threshold, i.e., almost everywhere.
In Sec.~\ref{sec:test} characteristic features of laminar chaos in systems with random and chaotically time-varying delay are identified and the test for laminar chaos introduced in \cite{muller-bender_laminar_2020,muller-bender_pseudolaminar_2023} is generalized and benchmarked.

\section{Random and chaotically time-varying delay}
\label{sec:delay_definition}

In this section, we define the notion of random and chaotically time-varying delays that will be explored in this manuscript.
The original theory on laminar chaos was derived for periodic delay variations, where a sinusoidal delay of the form
\begin{equation}
    \label{eq:sindelay}
    \tau(t) = \tau_0 + \frac{A}{2\pi}\,\sin(2\pi\,t)
\end{equation}
served as a generic example.
This parameterization with the mean delay $\tau_0$ and the amplitude $A$ when used in Eq.~\eqref{eq:sys} enables a systematic analysis of observables in the delay parameter space formed by $A$ and $\tau_0$, revealing interesting features such as a huge variation of the attractor dimension as a function of $\tau_0$ in a fractal manner \cite{muller_laminar_2018}.
We replace the sinusoidal delay with random and chaotically varying functions $\tau(t)=\tau_{\omega}(t)$, while keeping this form of parameterization.
The index $\omega$ represents the disorder parameter and therefore determines the specific realization of the random delay or the set of initial conditions of the chaotic delay generating process.
As done in Eq.~\eqref{eq:sindelay} the delay variation is normalized such that the amplitude $A$ can be varied between $0$ and $1$, where it is ensured that we always have 
\begin{equation}
    \label{eq:delaycondition}
    \dot{\tau}(t)<1, \text{ for almost all } t.
\end{equation}
This avoids several mathematical problems and can also be motivated by physical arguments \cite{verriest_inconsistencies_2011,verriest_state_2012}.

\begin{figure}
	\includegraphics[width=0.5\textwidth]{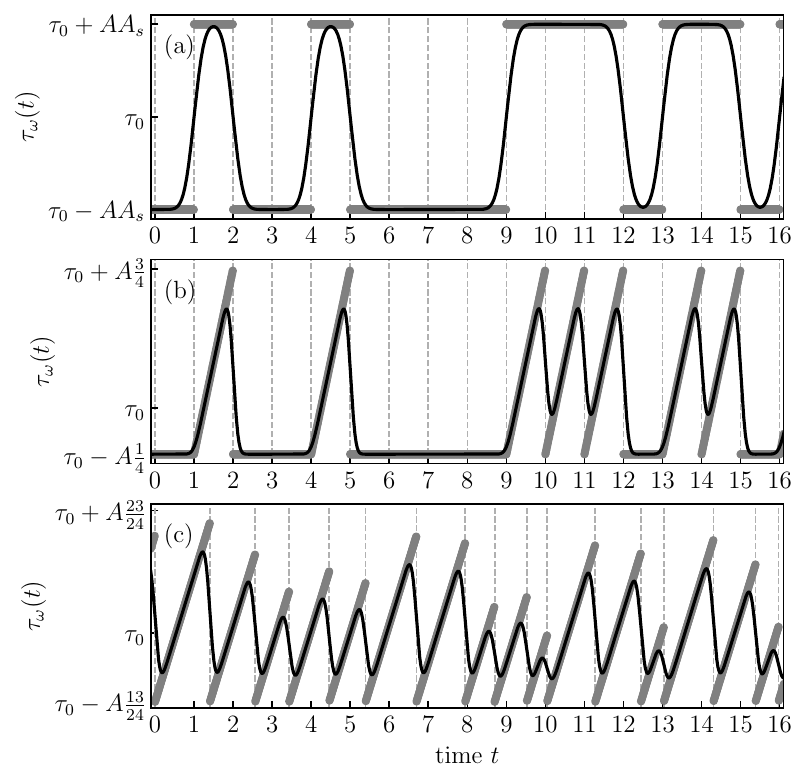}
	\caption{Three types of random delays (thin black lines), which are obtained by smoothing randomly generated function segments (thick gray lines) with a Gaussian kernel with standard deviation $\varsigma$:
    (a) dichotomic noise with a dwell time equal to one, $\varsigma=0.2$; (b) sawtooth wave that is randomly switched on and off, $\varsigma=0.1$; (c) sawtooth wave that increases with constant slope for a randomly chosen time and then resets, $\varsigma=0.1$. 
	}
	\label{fig:randomdelay}
\end{figure}

\begin{figure}
	\includegraphics[width=0.5\textwidth]{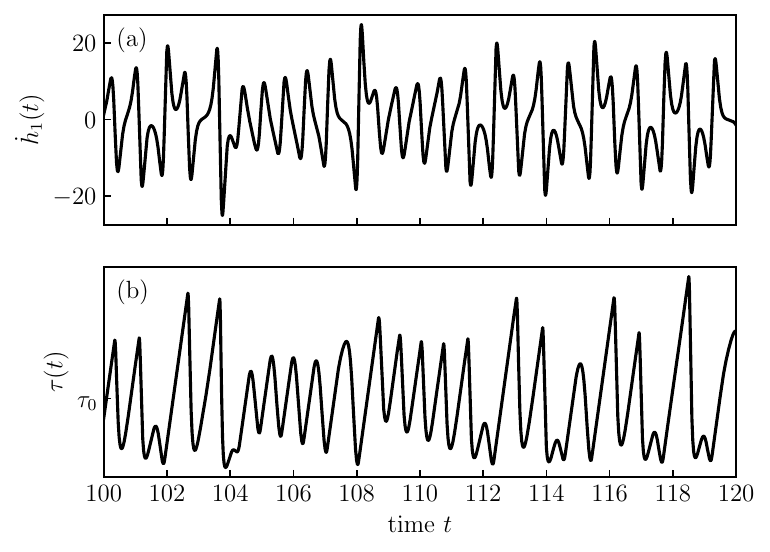}
	\caption{Example of a chaotically time-varying delay:
        The derivative $\dot{h}_1(t)$ of the first component of the Lorenz system, Eqs.~(\ref{eq:lorenz_1}-\ref{eq:lorenz_3}), shown in (a) is filtered by Eq.~\eqref{eq:lorenz_filter}.
        The resulting signal $h_4(t)$ is shifted and scaled according to Eq.~\eqref{eq:chaotic_delay} leading to the time-varying delay $\tau(t)$ shown in (b).
	}
	\label{fig:chaoticdelay}
\end{figure}

For our numerical analysis, we consider three examples of generic smooth random functions, which are illustrated in Fig.~\ref{fig:randomdelay}.
The differences between them allow us to distinguish between general results and more specific results that follow from the structure of the delay variation.
Each random delay $\tau_{\omega}(t)$ is smoothed, scaled and shifted version of a piecewise linear signal $\chi_{\omega }(t)$, which determines the statistics
\footnote{In principle, a random delay can be defined as the limit of a quasiperiodic delay with an infinite number of randomly chosen frequencies leading to first interesting results \cite{muller-bender_laminar_2023}.
However, to fulfill Eq.~\eqref{eq:delaycondition} in the generic case of a sum of cosines with random frequencies as considered in \cite{muller-bender_laminar_2023}, the overall amplitude of the delay tends to zero as the number of frequencies increases such that the limit of an infinite number of frequencies leads to a constant delay.
We therefore use a different approach.}.
We choose three different signals $\chi_{\omega }(t)$:

\paragraph{Dichotomic delay}
$\chi_{\omega }(t)$ is a piecewise constant dichotomic signal with values $S_{i}=\pm 1$ and a residence time equal to one, i.e., we have
\begin{equation}
    \label{eq:randomdelay_dichotomic}
    \chi_{\omega }(t) = S_{i} \text{ for } t\in [i-1,i),
\end{equation}
where $i$ are integers.
\paragraph{On-off sawtooth delay}
$\chi_{\omega }(t)$ is given by a periodic sawtooth wave, which is randomly switched on and off.
We define it for $t\in [i-1,i)$ as
\begin{equation}
    \label{eq:randomdelay_onoffsaw}
    \chi_{\omega }(t) = (t \text{ mod } 1)\, S_{i} - c, \quad S_{i} \in \{0, 1\},
\end{equation}
where the constant $c$ is chosen such that the time average of $\chi_{\omega }(t)$ vanishes for almost all realizations.

In cases (a) and (b), a realization is determined by an infinite symbol chain $\omega=S_{1}S_{2}\ldots$ of independent random symbols $S_{i}$, which are distributed with equal probability.

\paragraph{Random-duration sawtooth}
Inspired by the sawtooth-shaped random delay in \cite{verriest_stability_2009}, $\chi_{\omega }(t)$ is given by a sawtooth wave with slope equal to one, where the $i$th linear segment lasts for a randomly chosen time $S_i$ and then, after resetting to a fixed initial value, the next linear segment with duration $S_{i+1}$ begins.
We have for $t \in [T_{i-1},T_i)$
\begin{equation}
    \label{eq:randomdelay_randdursaw}
    \chi_{\omega }(t) = t - T_{i-1} - c, \text{ with } T_i = \sum_{j=1}^{i} S_j \text{ for } i>0
\end{equation}
and $T_0=0$, where the constant $c$ is chosen such that the time average of $\chi_{\omega }(t)$ vanishes for almost all realizations.
The $S_i$ are independent and uniformly distributed in $[1/2,3/2]$, where the mean duration of the linear segments equals the duration of the linear segments in the on-off sawtooth delay.

In all cases, the random delay variation is then obtained by a convolution $(G_{\varsigma }\ast \chi_{\omega})(t)$ with a Gaussian $G_{\varsigma}(t)$ of variance $\varsigma^2$.
Scaling by $AA_{s}$ and shifting by an offset $\tau_0$ gives our random delay
\begin{equation}
	\label{eq:random_delay}
	\tau_{\omega}(t)=\tau_{0}+AA_{s} \, (G_{\varsigma}\ast \chi_{\omega})(t),
\end{equation}
where $A_{s}=\sqrt{2\pi\varsigma^{2}}/2$ for the dichotomic delay, Fig.~\ref{fig:randomdelay}(a), and $A_{s}=1$ for both the on-off sawtooth and random duration sawtooth delay, Fig.~\ref{fig:randomdelay}(b,c), guarantees that Eq.~\eqref{eq:delaycondition} holds for $A\in [0,1]$.

As a generic example of a chaotically time-varying delay we consider
\begin{equation}
    \label{eq:chaotic_delay}
    \tau_{\omega}(t)=\tau(\vec{h}(t))=\tau_0+A\,(h_4(t)-\bar{h}_4),
\end{equation}
where the delay generating process is given by
\begin{subequations}
	\label{eq:lorenz}
	\begin{align}
		\dot{h}_1 &= \sigma (h_2 - h_1) \label{eq:lorenz_1}\\
		\dot{h}_2 &= h_1(\rho - h_3) - h_2 \label{eq:lorenz_2}\\
		\dot{h}_3 &= h_1 h_2 - \beta\,h_3 \label{eq:lorenz_3}\\
		\dot{h}_4 &= 1 - h_4\, \ln \left(1+e^{-\gamma\,\dot{h}_1} \right) \label{eq:lorenz_filter}.
	\end{align}
\end{subequations}
The first three equations make up the Lorenz system \cite{lorenz_deterministic_1963}, where we use the parameters $\sigma=10$, $\rho=28$, $\beta=\frac{8}{3}$ for chaos as used originally by Lorenz.
The constant $\bar{h}_4=\lim_{T\to\infty} \frac{1}{T} \int_{t_0}^{t_0+T} dt\, h_4(t)$ ensures that the time average vanishes.
The delay variation is given by the solution of Eq.~\eqref{eq:lorenz_filter}, which is a filtered version of the derivative $\dot{h}_1$ of the first component of the Lorenz system.
The nonlinear filter is designed such that the delay variation mimics the random duration sawtooth delay shown in Fig.~\ref{fig:randomdelay}(c)
\footnote{The sawtooth shape appears since we have $\dot{h}_4 \approx 1$ for $\gamma \dot{h}_1 \gg 0$ and $\dot{h}_4 \approx 1 +\gamma\, \dot{h}_1\,h_4$ for $\gamma \dot{h}_1 \ll 0$.
	This means that $h_4(t)$ increases approximately linearly with slope $1$ if $\gamma \dot{h}_1 \gg 0$ and rapidly decreases if $\gamma \dot{h}_1 \ll 0$.}, since for sawtooth-shaped variations the strongest effect of the time-varying delay is observed as we will in Sec.~\ref{sec:numerics} and Sec.~\ref{sec:theory}.
However, we note that the theory presented here is not restricted to such variations and is valid for all short time correlated chaotically varying delays. 
In the following numerical analysis we choose the filter parameter $\gamma=0.2$, which leads to the chaotically time-varying delay shown in Fig.~\ref{fig:chaoticdelay}(b), where the unfiltered chaotic signal $\dot{h}_1(t)$ is shown in Fig.~\ref{fig:chaoticdelay}(a).

\section{Review of laminar chaos}
\label{sec:review}

In this section, we briefly recall the key mechanism behind laminar chaos, which was first found for Eq.~\eqref{eq:sys} with periodic delay \cite{muller_laminar_2018}, $\tau(\vec{h}(t))=\tau(t)=\tau(t+1)$, and later generalized to quasiperiodic delays \cite{muller-bender_laminar_2023}.
In principle, the system defined by Eq.~\eqref{eq:sys} is a feedback loop, where a signal is delayed and frequency modulated by a time-varying delay $\tau(t)$.
The intensity of the signal is transformed by the nonlinearity $f$ and after that the signal is low-pass filtered with a cutoff frequency $\Theta$ by the first order filter defined by the left hand side of Eq.~\eqref{eq:sys} before the next roundtrip inside the feedback loop begins.
This structure is reflected in the iterative solution of the system using the \emph{method of steps} \cite{bellman_computational_1961,bellman_computational_1965}, where the solution $z(t)$ is divided into solution segments $z_k(t)=z(t)$ with $t \in \mathcal{I}_k=(t_{k-1},t_k]=(t_k-\tau(t_k),t_k]$.
The domains of the $z_k(t)$ are the so-called state-intervals and their interval boundaries $t_k$ are connected by the so-called access map given by
\begin{equation}
	\label{eq:access_map}
	t'=R(t) = t-\tau(t).
\end{equation}
Given that Eq.~\eqref{eq:delaycondition} is fulfilled, the access map is monotonically increasing and therefore the state-intervals do not intersect and their union covers the whole domain of the solution $z(t)$.
Now we replace $\dot{z}(t)$ and $z(t)$ with $\dot{z}_{k+1}(t)$ and $z_{k+1}(t)$, respectively, on the left hand side of Eq.~\eqref{eq:sys} and $z(t-\tau(t))=z(R(t))$ with $z_k(R(t))$ on the right hand side.
Solving the resulting equation for $z_{k+1}(t)$ gives
\begin{equation}
	\label{eq:soluop}
	z_{k+1}(t) = z_{k}(t_{k}) e^{-\Theta(t-t_{k})} + \int\limits_{t_{k}}^{t} \! dt' \, \Theta e^{-\Theta(t-t')} f(z_k\bm{(}R(t')\bm{)}),
\end{equation}
a (nonlinear) solution operator generating the solution segment $z_{k+1}(t)$ from the preceding solution segment $z_k(t)$.
The low pass filter acts as a smoothing operator with exponential kernel $\Theta e^{-\Theta(t-t')}$ of width $(1/\Theta)$.
We focus on the limit of a large cutoff frequency $\Theta$ and therefore we consider $\Theta\to\infty$, where the kernel converges to a delta distribution so that the smoothing operator becomes the identity \cite{ikeda_high-dimensional_1987}.
We then obtain the so-called limit map given by
\begin{equation}
	\label{eq:limit_map}
	z_{k+1}(t) = f(z_k\bm{(}R(t)\bm{)}),
\end{equation}
which is also a good approximation of Eq.~\eqref{eq:soluop} for finite $\Theta$ given that $\dot{z}(t) \ll \Theta$, i.e, when the timescale of the solution $z(t)$ is much larger than the width $(1/\Theta)$ of the kernel.
Under the limit map, the evolution of the segment $z_k(t)$ can be interpreted as iteration of the graph $(t,z_k(t))$ by the two dimensional map
\begin{subequations}
	\label{eq:2d_map}
	\begin{align}
		x_k &= R^{-1}(x_{k-1}) \label{eq:2d_map_r}\\
		y_k &= f(y_{k-1}) \label{eq:2d_map_f},
	\end{align}
\end{subequations}
which consists of two independent one-dimensional maps.
Equation~\eqref{eq:2d_map_r} is the inverse of the access map given by Eq.~\eqref{eq:access_map} and Eq.~\eqref{eq:2d_map_f} is a map defined by the nonlinearity $f$ of the feedback.
For periodic or quasiperiodic delay, the access map is the lift of a circle map (cf. \cite{katok_introduction_1997}) or a foliation preserving torus map (cf. \cite{petrov_torus_2003,he_resonances_2023}), respectively.
For general systems with (quasi-)periodic delay it is known that the dynamics of these maps define two universal classes of time-varying delays \cite{otto_universal_2017,muller_dynamical_2017,muller-bender_laminar_2023}.
If the dynamics is marginally stable dynamics with vanishing Lyapunov exponent, $\lambda[R]=0$, and perturbations of a reference orbit do on average neither grow nor decay, the delay belongs to the class of conservative delays.
Such delays are equivalent to constant delays in the sense that the delay system can be transformed into a system with constant delay but time-dependent coefficients.
This means that they show qualitatively the same dynamics.
In contrast, dissipative delays belong to stable access map dynamics, $\lambda[R]<0$.
Such delays are not equivalent to constant delay systems, and thus they can lead to drastic differences in the dynamics compared to constant delay systems.

These differences become very pronounced for chaotic dynamics of Eq.~\eqref{eq:sys} at large $\Theta$.
Chaos here is weak in the sense of \cite{heiligenthal_strong_2011} (not to be confused with weak chaos from chaotic diffusion \cite{albers_weak_2024}) as it originates from the delayed feedback due to the negative instantaneous Lyapunov exponent.
Given that the map defined by the feedback nonlinearity, Eq.~\eqref{eq:2d_map_f}, is chaotic, stretching and folding of the function values of the signal leads to strong fluctuations as the signal circles inside the feedback loop.
For conservative delays, the delay variation induces only an additional frequency modulation due to the Doppler effect, which can not compensate the stretching and folding induced by the feedback nonlinearity.
Therefore the characteristic frequency of the signal increases with each roundtrip inside the feedback loop and is only bounded by the cut-off frequency $\Theta$ of the low-pass filter.
This is the mechanism behind \emph{turbulent chaos} \cite{ikeda_high-dimensional_1987}, also found in constant delay systems and named after optical turbulence in an optical ring cavity \cite{ikeda_optical_1980}.
It is a very high-dimensional dynamics as the dimension of the chaotic attractor is proportional to $\Theta$ and thus can be arbitrarily large \cite{farmer_chaotic_1982}.
In stark contrast, dissipative delays can lead to comparably low-dimensional chaos as demonstrated in \cite{muller_laminar_2018,muller-bender_resonant_2019,muller-bender_laminar_2023}.
The main mechanism behind this is the so-called resonant Doppler effect, where \emph{the frequency modulation due to the delay variation is in resonance with the average roundtrip time inside the feedback loop} - a consequence of the mode-locking behavior of the access map.
The competition between the resonant Doppler effect and the stretching and folding of the feedback nonlinearity leads to the development of low frequency phases that are periodically or quasiperiodically interrupted by rather short high frequency phases, which characterize the comparably low-dimensional types of chaos called (generalized) laminar chaos.
Laminar chaos occurs when
\begin{equation}
	\label{eq:lamchaos_crit}
	\lambda[f] + \lambda[R] < 0,
\end{equation}
where $\lambda[f]$ is the Lyapunov exponent of the map defined by Eq.~\eqref{eq:2d_map_f}.
Then the low frequency phases degenerate to nearly constant laminar phases provided their duration is much larger than the width $(1/\Theta)$ of the smoothing kernel in Eq.~\eqref{eq:soluop}.
Since the derivative of these laminar phases is small, the two-dimensional map, Eq.~\eqref{eq:2d_map} well approximates their dynamics, where \eqref{eq:2d_map_r} and \eqref{eq:2d_map_f} governs the dynamics of the durations and the levels of the laminar phases, respectively.
Neglecting the high-frequency phases, which degenerate to short burstlike transitions between the laminar phases, laminar chaos can be described by this low-dimensional system leading to a very low-dimensional chaotic attractor.
If $\lambda[R]$ is negative but Eq.~\eqref{eq:lamchaos_crit} is not fulfilled, generalized laminar chaos of order $(m-1)$ is observed, where $m>0$ is the smallest integer such that
\begin{equation}
	\label{eq:genlamchaos_crit}
	\lambda[f] + m\,\lambda[R] < 0
\end{equation}
holds.
In this case, $(m-1)$-order polynomials take the role of the nearly constant phases of classical laminar chaos leading to an effective dimension of order $O(m)$, which is small if $m$ is small \cite{muller-bender_resonant_2019}.

In this paper, we generalize this theory to systems with random and chaotically time-varying delay.
Using the theory of disordered circle maps provided in \cite{muller-bender_suppression_2022}, we demonstrate that laminar chaos can be observed in such systems and the condition for laminar chaos, Eq.~\eqref{eq:lamchaos_crit}, remains valid.
We use that access maps for random delay can be treated as infinite size limit, $L\to\infty$, of disordered circle maps with size and period $L$ given by
\begin{equation}
	\label{eq:circle_map}
	\theta' = R(\theta) \mod L,
\end{equation}
where the delay $\tau(t)$ is defined as a random function in an interval of length $L$ and repeated periodically.
So for all finite sizes $L$, we can rely on the known properties of periodic delays, while approaching the infinite size limit $L\to\infty$.
In this limit, almost all realizations of the random delay lead to the same well-defined Lyapunov exponent $\lambda[R]$.
Hence Eqs.~(\ref{eq:lamchaos_crit},\ref{eq:genlamchaos_crit}) apply directly, as they follow from a stability analysis of polynomial solutions of the limit map, Eq.~\eqref{eq:limit_map}, only assuming that the maps defined by $f$ and $R$ have a well defined Lyapunov exponent (see \cite{muller_laminar_2018,muller-bender_resonant_2019} for details).
While randomness of the delay variation does not affect the validity of the criteria for (generalized) laminar chaos, it drastically changes the delay parameter space: These low-dimensional types of chaos occupy almost all of it, such that high-dimensional turbulent chaos becomes extremely rare. 
In Sec.~\ref{sec:test} we also find interesting changes in the dynamical properties of laminar chaos compared to periodic and quasiperiodic delay systems, which make it necessary to update the time series analysis toolbox for laminar chaos developed in \cite{hart_laminar_2019,muller-bender_laminar_2020,muller-bender_pseudolaminar_2023}. 

\section{Laminar chaos in systems with random and chaotically time-varying delay}
\label{sec:lamchaos_rand}

In this section, we demonstrate that laminar chaos and its generalizations can be ubiquitously observed in systems with random and chaotically time-varying delay.
We first present numerical results in Sec.~\ref{sec:numerics}, where we consider Eq.~\eqref{eq:sys} with the time-varying delays defined in Sec.~\ref{sec:delay_definition}.
By scanning the delay parameter space spanned by the mean delay $\tau_0$ and the delay amplitude $A$, we find that low-dimensional (generalized) laminar chaos dominates almost the whole parameter space, while high-dimensional turbulent chaos is suppressed for most parameters.
This is in stark contrast to periodic and quasiperiodic delay variations, where the parameter sets for both types of chaos typically have a nonzero measure.
The numerical results are explained by theoretical considerations in Sec.~\ref{sec:theory}.

\subsection{Numerical results}
\label{sec:numerics}

\begin{figure*}
	\includegraphics[width=0.99\textwidth]{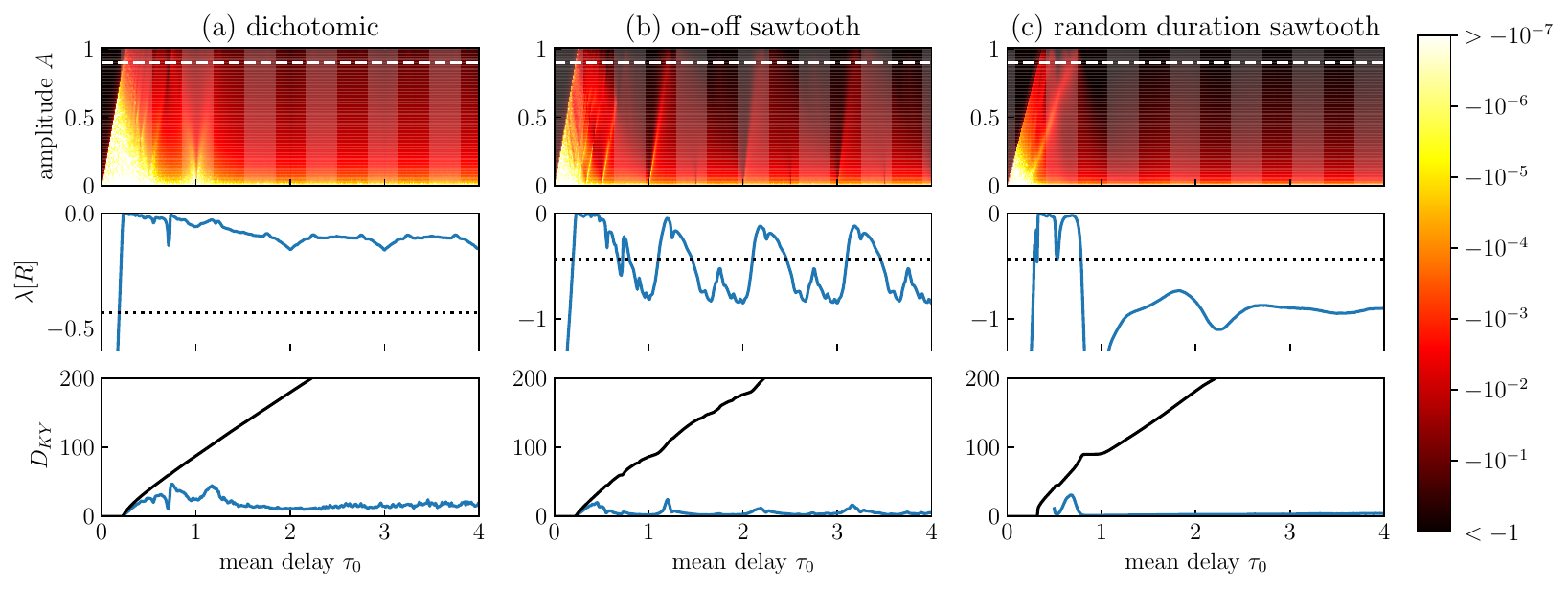}
	\caption{Influence of the random delays shown in Fig.~\ref{fig:randomdelay} on the dimension of chaotic attractors:
        In the top panel a heat map of Lyapunov exponent $\lambda[R]$ of the access map, Eq.~\eqref{eq:access_map},  is shown as a function of the mean delay $\tau_0$ and the delay amplitude $A$.
        For fixed $A=0.9$ (white dashed line), $\lambda[R]$ and the Kaplan-Yorke dimension $D_{KY}$ of Eq.~\eqref{eq:sys} is shown in the center and bottom panels (blue solid lines), respectively.
        While for smaller $\tau_0$ we find fractal structures similar to those in Fig.~\ref{fig:arnoldkydim_sin}, for increasing $\tau_0$ the parameter space becomes more and more regular, where only dissipative delays, $\lambda[R]<0$ appear.
        Therefore, in almost the whole parameter space, a drastic reduction of $D_{KY}$ compared to a system with constant delay (black solid line, see text) is observed.
        Laminar chaos is found for $\lambda[f]+\lambda[R]<0$, i.e., whenever $\lambda[R]$ is below the dotted line in the center panels.
	}
	\label{fig:arnoldkydim}
\end{figure*}

\begin{figure}
	\includegraphics[width=0.5\textwidth]{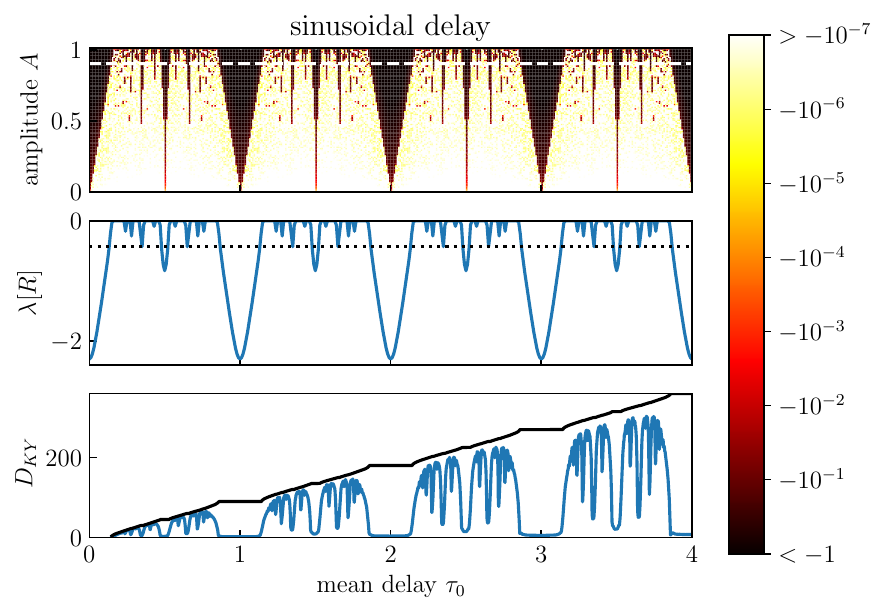}
	\caption{Same analysis as in Fig.~\ref{fig:arnoldkydim} for a sinusoidally varying delay:
        The fractal structures of the delay classes revealed by the heat map of $\lambda[R]$ (top panel) disappear for random and chaotically time-varying delays if $\tau_0$ is large enough.
        Conservative delays, $\lambda[R]=0$, lead to high-dimensional turbulent chaos, where the Kaplan-Yorke dimension $D_{KY}$ (bottom panel, blue line) is close to that of a comparable constant delay system (black solid line, see text).
        In contrast, for dissipative delays, $\lambda[R]<0$, a drastic reduction of $D_{KY}$ is observed leading to a variation of $D_{KY}$ over several orders of magnitude, which disappears for random and chaotically time-varying delays.
	}
	\label{fig:arnoldkydim_sin}
\end{figure}

\begin{figure}
	\includegraphics[width=0.5\textwidth]{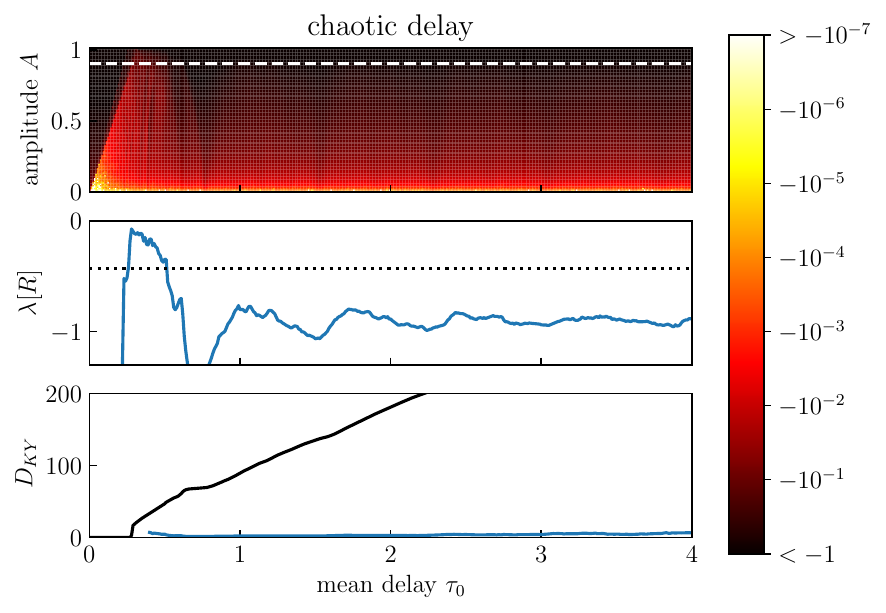}
	\caption{Same analysis as in Fig.~\ref{fig:arnoldkydim} for the chaotically time-varying delay shown in Fig.~\ref{fig:chaoticdelay}:
        The results are qualitatively equivalent to those we obtained for random delays.
	}
	\label{fig:arnoldkydim_chaotic}
\end{figure}

In the following, we consider Eq.~\eqref{eq:sys} with a quadratic nonlinearity $f(z)=3.8\,z(1-z)$ and the random and chaotically time-varying delays illustrated in Fig.~\ref{fig:randomdelay} and  Fig.~\ref{fig:chaoticdelay}(b), respectively.
For the numerical analysis we choose $\Theta=200$ such that the small time scale $(1/\Theta)$ of the system is much smaller than the delay $\tau(t)$ and also much smaller than the time scale of the temporal structures of the delay variation.
The results are presented in Figs.~\ref{fig:arnoldkydim} and \ref{fig:arnoldkydim_chaotic} for random and chaotically time-varying delays, respectively.
For highlighting the differences between these and regular delay variations, analogous results for a periodic (sinusoidal) delay variation are shown in Fig.~\ref{fig:arnoldkydim_sin}.
Since laminar chaos and its generalization can only be observed in systems with dissipative delay, our first step is the classification of the time-varying delays in the delay parameter space using the Lyapunov exponent $\lambda[R]$ of the access map.
For general one-dimensional maps, the Lyapunov exponent is defined by \cite{ott_chaos_2002}
\begin{equation}
	\label{eq:lyapunov_exponent}
	\lambda[R] = \lim_{K\to\infty} \frac{1}{K} \sum_{k=0}^{K-1} \ln|\dot{R}\bm{(} R^k(t_0) \bm{)}|,
\end{equation}
where $R^k(t)=R\bm{(} R^{k-1}(t) \bm{)}$ is the $k$th iteration of $R(t)$.
A negative Lyapunov exponent, $\lambda[R]<0$, implies that the delay is a dissipative delay, whereas delays with $\lambda[R]=0$ are potential candidates for conservative delays
\footnote{Here we skip the additional tests needed for the exact classification in the case $\lambda[R]=0$ since dissipative delays with $\lambda[R]=0$ are typically of measure zero with respect to the delay parameter space.
For periodic delays, this happens at the boundaries of the so-called Arnold tongues, where the circle map given by Eq.~\eqref{eq:circle_map} shows a tangential bifurcation.   
In principle, dissipative delays with $\lambda[R]=0$ would lead to the limiting case of generalized laminar chaos with order $m\to\infty$ (see Eq.~\eqref{eq:genlamchaos_crit}).}.
Heat maps of the $\lambda[R]$ as a function of $\tau_0$ and $A$ can be found in the top panels of Figs.~\ref{fig:arnoldkydim}-\ref{fig:arnoldkydim_chaotic}
\footnote{For each parameter pair, the Lyapunov exponent $\lambda[R]$ was computed using a finite number of iterations $K$, where we used $K=10^6$ for the sinusoidal delay, $K=5\cdot 10^5$ for the random delays, and $K=5\cdot 10^4$ for the chaotically time-varying delay. To let the transients relax, the initial condition $t_0$ was iterated at least $10^3$ times prior to the computation of $\lambda[R]$.}.
Since disspative delays lead to (generalized) laminar chaos, this gives us an overview of the distribution of these types of dynamics in the delay parameter space.
As a next step, we fix the delay amplitude $A=0.9$ and plot $\lambda[R]$ as a function of the mean delay $\tau_0$ to check, where the condition for classical laminar chaos, Eq.~\eqref{eq:lamchaos_crit}, is fulfilled.
The results are found in the center panels of Figs.~\ref{fig:arnoldkydim}-\ref{fig:arnoldkydim_chaotic}, where laminar chaos is expected if the blue solid line, which represents $\lambda[R]$, is below the dotted line, which corresponds to $\lambda[f]+\lambda[R]=0$.
Finally, we consider the delay system with the considered delays and compute the Kaplan-Yorke dimension $D_{KY}$ defined by \cite{kaplan_chaotic_1979,farmer_dimension_1983}
\begin{equation}
	D_{KY} = j + \frac{\sum_{m=0}^{j-1} \lambda_m}{|\lambda_j|},
\end{equation}
where $j$ is the largest integer such that $\sum_{m=0}^{j-1} \lambda_m \geq 0$.
In the bottom panels of Figs.~\ref{fig:arnoldkydim}-\ref{fig:arnoldkydim_chaotic}, $D_{KY}$ is shown as a function of $\tau_0$ for fixed $A=0.9$ (blue line)
\footnote{For the estimation of $D_{KY}$, the Lyapunov exponents $\lambda_m$ of the delay system, Eq.~\eqref{eq:sys}, were computed using the method in \cite{farmer_chaotic_1982}. The involved delay differential equations are solved via the two-stage Lobatto IIIC method with linear interpolation \cite{bellen_numerical_2003} using integration steps of width $\Delta t=2\cdot 10^{-5}$ and $\Delta t=2\cdot 10^{-3}$ for the nonlinear and the linearized system, respectively. The separation functions governed by the linearized system were reorthonormalized after each time units, i.e., after each $500$ integration steps. After an initialization period of $5000$ time units for the separation functions, the Lyapunov exponents $\lambda_m$ were computed for the $5000$ following time units.},
such that we can relate the dimensionality of the chaotic dynamics to the Lyapunov exponent $\lambda[R]$ of the access map shown in the center panels, which shows us, whether the condition for laminar chaos is fulfilled.
In addition, an estimate for the Kaplan-Yorke dimension of a comparison system with constant delay, $\tau(t)=T$, is shown (black line), where the constant delay is set to the average roundtrip time $T=v[R^{-1}]=-v[R]$ inside the feedback loop, which is equal to the negative drift velocity $v[R]$ of the access map defined by
\begin{equation}
	\label{eq:drift_velocity}
	v[R] = \lim_{k\to\infty} \frac{R^k(t_0) - t_0}{k}.
\end{equation}
This gives us an upper bound for the Kaplan-Yorke dimension $D_{KY}$ for time-varying delay, which is approximately reached for turbulent chaos (see local maxima of $D_{KY}$ in Fig.~\ref{fig:arnoldkydim_sin})
\footnote{Following the numerical results in \cite{muller-bender_laminar_2023}, $D_{KY} \approx 90\,T$ is a good estimate for the system with the parameters considered here, where the constant delay is set to $T$.
There it also was found that this estimate is an upper bound for conservative time-varying delays, since then the system can be transformed to a constant delay system with time-varying coefficients, which only lead to a slight decrease of the $D_{KY}$ compared to a system with constant coefficients}.

To highlight the differences that come with random and chaotically time-varying delay, we first consider a generic example of a regular delay variation: The sinusoidal delay given by Eq.~\eqref{eq:sindelay}.
Its parameter space shown in the top panel of Fig.~\ref{fig:arnoldkydim_sin} is characterized by two dynamical regimes of the access map and its corresponding circle map, Eq.~\eqref{eq:circle_map} with $L=1$.
Inside the so-called Arnold tongues \cite{arnold_small_1961,*arnold_small_1961_erratum}, which are the dark tongue-like structures where $\lambda[R]$ is negative, the circle map shows stable periodic dynamics and the resulting delays are classified as dissipative delays.
The complement of the Arnold tongues, which for fixed $A$ is a Cantor set with nonzero measure, i.e., a fat fractal \cite{ott_chaos_2002}, leads to marginally stable quasiperiodic dynamics of the circle map, $\lambda[R]=0$ and thus the resulting delay are classified as conservative delays.
As illustrated in the center panel, $\lambda[R]$ as a function of $\tau_0$ for fixed $A=0.9$ also reflects this fractal structure showing an irregular behavior while it is periodic in $\tau_0$, where the period equals the delay period.
For the delay system this means that the dynamics changes irregularly between high-dimensional turbulent chaos, and low-dimensional (generalized) laminar chaos if the delay parameters are varied.
This has drastic consequences for the Kaplan-Yorke dimension $D_{KY}$ as shown in the bottom panel for fixed $A=0.9$.
If the mean delay $\tau_0$ is varied, one observes huge variations of $D_{KY}$ over several orders of magnitudes, where the local maxima are roughly proportional to $\tau_0$.
The overall variation of $D_{KY}$ increases with the parameter $\Theta$ since the local maxima are also proportional to $\Theta$ as they correspond to conservative delays leading to turbulent chaos, where the system is equivalent to a constant delay system so that the theory in \cite{farmer_chaotic_1982} applies.
In contrast, the local minima hardly vary with $\Theta$ given that $\Theta$ is large enough since the system is then well described by the low-dimensional dynamics of the limit map, Eq.~\eqref{eq:limit_map}.
Such huge variations of $D_{KY}$ in a fractal manner are also observed for quasiperiodic delays \cite{muller-bender_laminar_2023}.

\begin{figure}
    \centering
    \includegraphics[width=0.5\textwidth]{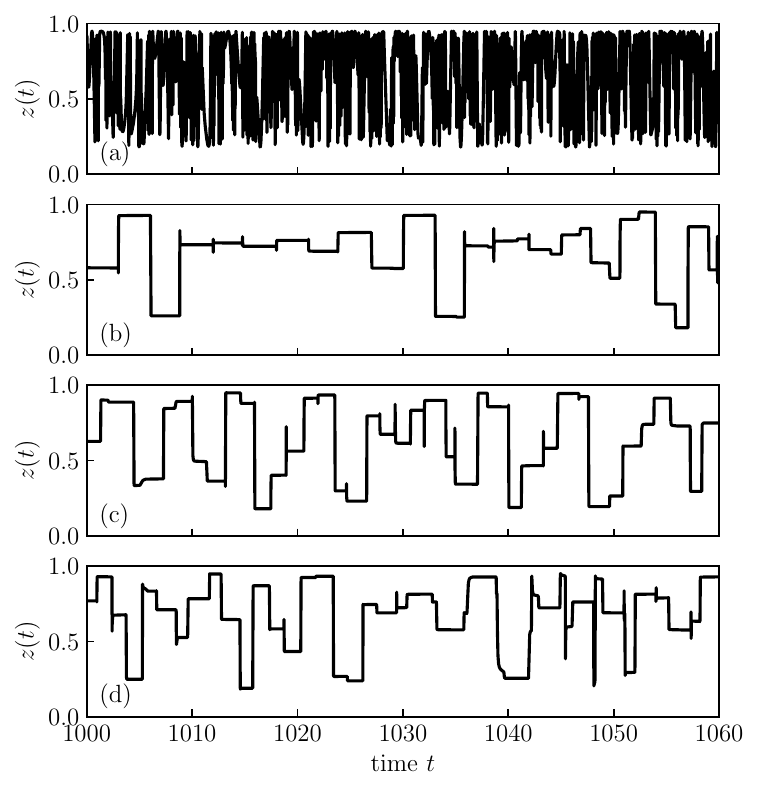}
    \caption{Time series of Eq.~\eqref{eq:sys} with the random delays as described in Fig.~\ref{fig:randomdelay}(b,c) and chaotically time-varying delay in Fig.~\ref{fig:chaoticdelay}(b), where we set the delay amplitude to $A=0.9$:
        Generalized laminar chaos is observed if we have $\lambda[R]<0$ and the condition for laminar chaos $\lambda[f]+\lambda[R]<0$ is not fulfilled, as shown in (a) for on-off sawtooth delay, Fig.~\ref{fig:randomdelay}(b), with $\tau_0=3.25$.
        Laminar chaos is observed for $\lambda[f]+\lambda[R]<0$ as shown in (b), (c), and (d) for on-off sawtooth delay, random duration sawtooth delay, and chaotically time-varying delay, respectively, where we set $\tau_0=3$.
    }
    \label{fig:trajectories}
\end{figure}

The situation drastically changes, when it comes to a random delay variation as shown in Fig.~\ref{fig:arnoldkydim}.
While for small values of the mean delay $\tau_0$ we still find fractal structures in the parameter space similar to the ones observed for periodic delays, the fractal structures disappear for larger $\tau_0$.
Moreover, the Lyapunov exponent of the access map seems strictly negative for all $A>0$ if $\tau_0$ exceeds a certain threshold, which indicates that almost all parameters lead to dissipative delays.
This is also confirmed in the center panels, where $\lambda[R]$ is shown as a function of $\tau_0$ for fixed $A=0.9$.
According to the theory on laminar chaos in Sec.~\ref{sec:review}, the delay system defined by Eq.~\eqref{eq:sys} shows low-dimensional (generalized) laminar chaos for almost all delay parameters, when $\tau_0$ exceeds a certain threshold.
In Sec.~\ref{sec:theory} we find that this threshold is of the order of the correlation length of the delay variation.
The low-dimensionality of the dynamics of the delay system is confirmed by the bottom panels, where we find that the Kaplan-Yorke dimension $D_{KY}$ for random delays (blue line) with large enough $\tau_0$ is of orders of magnitudes smaller than the dimension observed for the comparison system with a constant delay (black line).
This was already conjectured based on results for a quasiperiodic delay with a large number of frequencies \cite{muller-bender_laminar_2023} and is now confirmed.

As shown in Fig.~\ref{fig:arnoldkydim_chaotic}, basically the same results are obtained for the considered chaotically time-varying delay.
From a qualitative point of view the structure of the parameter space is most similar to the sawtooth-shaped delay with random duration, see Fig.~\ref{fig:randomdelay}(c) and Fig.~\ref{fig:arnoldkydim}(c).
This is not surprising as the main difference given by the periodic structures of the parameter space for the delay variations considered in Fig.~\ref{fig:arnoldkydim}(a,b) are caused by an inherent periodicity of the delay generating process as shown in Sec.~\ref{sec:theory}.
For the sawtooth-shaped delay with random duration and the chaotic delay there is no such inherent periodicity.
The apparent periodicity in the top panel of Fig.~\ref{fig:arnoldkydim_chaotic} given by the periodic occurrence of lighter and darker regions is caused by the structure of the correlation function of the Lorenz system, which periodically oscillates while converging to zero (see \cite{grossmann_correlation_1982}).
Therefore the apparent periodic structure vanishes for $\tau_0\to\infty$ as indicated by $\lambda[R](\tau_0)$ in the center panel.

In Fig.~\ref{fig:trajectories}, exemplary time series of the delay system with the considered random and chaotically time-varying delays are shown, where we choose the delay amplitude $A=0.9$ in accordance with the analysis above.
For this amplitude, the on-off sawtooth delay can lead to both laminar and generalized laminar chaos as shown in the center panel of Fig.~\ref{fig:arnoldkydim}(b), where the condition for laminar chaos, Eq.~\eqref{eq:lamchaos_crit} is fulfilled below the dotted line.
Time series of generalized laminar chaos and laminar chaos for that type of delay variation are shown in Fig.~\ref{fig:trajectories}(a,b), respectively.
The random duration sawtooth delay and the chaotically time-varying delay both lead to laminar chaos for large enough $\tau_0$ as indicated in the center panels of Fig.~\ref{fig:arnoldkydim}(c) and Fig.~\ref{fig:arnoldkydim_chaotic}.
Exemplary time series for these cases are show in Fig.~\ref{fig:trajectories}(c,d).
For the dichotomic delay and the chosen nonlinearity $f$ of the feedback, no laminar chaos is observed since the condition for laminar chaos is not fulfilled for any $\tau_0$ as shown in the center panel of Fig.~\ref{fig:arnoldkydim}(a).

\subsection{Theory}
\label{sec:theory}

In order to understand and verify the above numerical results, we provide analytical arguments using the theory of circle maps with quenched disorder from \cite{muller-bender_suppression_2022}.
According to the Sec.~\ref{sec:review}, the mechanism behind laminar chaos and generalized laminar chaos basically is governed by a competition between two one-dimensional maps, Eq.~\eqref{eq:2d_map}, which is expressed in terms of equations by the condition for laminar chaos, Eq.~\eqref{eq:lamchaos_crit}, and by the condition for generalized laminar chaos, Eq.~\eqref{eq:genlamchaos_crit}.
These conditions depend only on two characteristic quantities: The Lyapunov exponent $\lambda[f]$ of the map defined by the nonlinearity $f$ of the feedback and the Lyapunov exponent $\lambda[R]$ of the access map.
While the feedback nonlinearity is independent of the delay and therefore $\lambda[f]$ is still well defined when we consider random and chaotically time-varying delays, the access map is now defined by a random function $R(t)=R_\omega(t)=t-\tau_\omega(t)$, which in principle leads to a different value of $\lambda[R]$ for each realization $\omega$.
Another problem potentially affecting $\lambda[R]$ is the unboundedness of the dynamics of the access map due to the drift in negative direction following from $R(t)<t$ for all $t$. For periodic or quasiperiodic delays this is resolved by projecting the access map dynamics onto the circle or onto a torus leading to the circle map or a foliation preserving torus map, respectively, which are equivalent finite-dimensional dynamical systems with bounded dynamics.
For random and chaotically time-varying delays this is not possible.
In order to deal with that, we consider the access map as infinite size limit of a disordered circle map \cite{muller-bender_suppression_2022} with size $L$ given by
\begin{equation}
	\label{eq:disordered_circle_map}
	\theta_{k+1}= R_{\omega}(\theta_{k})\; \text{mod}\; L=[\theta_{k}-\tau_{\omega}(\theta_{k})]\; \text{mod}\; L.
\end{equation}
As done in Ref.~\cite{muller-bender_suppression_2022}, we can then rely on the well understood properties of the circle map.
For the random delays shown in Fig.~\ref{fig:randomdelay}(a,b) a system of size $L$ means that $L$ symbols $S_i$ are chosen and are repeated periodically to generate a periodic unsmoothed delay variation $\chi_\omega(t)$ with period $L$ from Eqs.~(\ref{eq:randomdelay_dichotomic},\ref{eq:randomdelay_onoffsaw}) before smoothing and shifting to finally obtain the time-varying delay.
The remaining random delay and the chaotically time-varying delay can be treated in a similar way.

Given that we have $\dot{\tau}(t)<1$ for almost all $t$, which is assumed for the whole analysis, the considered circle maps are so-called \emph{diffeomorphisms of the circle} \cite{katok_introduction_1997}.
These well understood dynamical systems show two types of dynamics, which can be classified by two characteristic quantities.
The first one is the rotation number given by
\begin{equation}
	\label{eq:rotationnumber}
	\rho_\omega = - \lim_{N\to\infty} \frac{1}{K\,L} \sum_{k=0}^{K-1} \tau_\omega(\theta_k).
\end{equation}
which is the average drift of an orbit $\{ \theta_k \}_{k\in\mathbb{N}_0}$ of the map, Eq.~\eqref{eq:disordered_circle_map}, divided by the period $L$.
For the delay system given by Eq.~\eqref{eq:sys}, it has a vivid interpretation as it is proportional to the average roundtrip time inside the feedback loop given by
\begin{equation}
    \label{eq:avroundtriptime}
    T_\omega = - L\, \rho_\omega,
\end{equation}
which is also equal to the average length of the state intervals $\mathcal{I}_k$ that were defined in Sec.~\ref{sec:review}.
The second quantity is the Lyapunov exponent $\lambda[R_\omega]$ as introduced before, which now can also be written as a time average over one orbit of the circle map, i.e.,
\begin{equation}
	\label{eq:lyapunovexponent_2}
	\lambda[R_\omega]=\lim_{K\to\infty}\frac{1}{K} \sum_{k=0}^{K-1} \ln(1-\dot{\tau}_\omega(\theta_k)).
\end{equation}
As found in standard literature such as \cite{katok_introduction_1997} such circle maps show two types of dynamics: Quasiperiodicity and mode-locking.
Marginally stable quasiperiodic dynamics is characterized by an irrational rotation number and $\lambda[R_\omega]=0$, whereas in the case of mode-locking the map has stable periodic orbits with period $q$ leading to a rational rotation number $\rho_\omega=-p/q$ and a negative Lyapunov exponent $\lambda[R_\omega]$.
In our case of circle maps with quenched disorder, where the disorder parameter $\omega$ denoting the specific realization is randomly chosen once and then stays constant as the circle map is iterated, the characteristic quantities are random variables, since they depend on the disorder parameter $\omega$.
However, in Ref.~\cite{muller-bender_suppression_2022} it is demonstrated that the disorder averages of $T_\omega$ and $\lambda[R_\omega]$ converge to well-defined limiting values for $L\to\infty$, where the variances vanishes asymptotically with $L^{-1}$ such that their distributions converge to delta distributions.
This means that almost all realizations lead to the same values of the average roundtrip time $T_\omega \to T$ and of the Lyapunov exponent $\lambda[R_\omega] \to \lambda[R]$ in this limit.
It follows that the theory of laminar chaos in Sec.~\ref{sec:review} can directly be applied to delay systems with random delays since almost all realizations of the random delay lead to the same dynamics of the access map so that the dynamics of the delay system can be uniquely classified by the condition for laminar chaos, Eq.~\eqref{eq:lamchaos_crit} and generalized laminar chaos, Eq.~\eqref{eq:genlamchaos_crit}.
This is also true for chaotically time-varying delays that are generated by an ergodic process (as it is the case for Eq.~\eqref{eq:chaotic_delay}, see \cite{tucker_lorenz_1999,luzzatto_lorenz_2005}) such that the results of the time averages in Eqs.~(\ref{eq:avroundtriptime},\ref{eq:lyapunovexponent_2}) are unique for almost all initial conditions $\omega$ in the limit $L\to\infty$.

The $\tau_0$-periodicity in the parameter space observed in Fig.~\ref{fig:arnoldkydim}(a,b) for larger $\tau_0$ is caused by an inherent $t$-periodicity of the derivative $\dot{\tau}_\omega(t)$ in a statistical sense.
While $\dot{\tau}_\omega(t)$ randomly varies in $t$, the probability distribution with respect to the disorder is periodic in $t$ due to the generation of the delay variation from randomly chosen function segments of constant length $1$, see Fig.~\ref{fig:randomdelay}(a,b), which leads to periodicity with period $1$ in the parameter space.

In order to explain the strictly negative value of the Lyapunov exponent of the access map that is observed in Fig.~\ref{fig:arnoldkydim} and Fig.~\ref{fig:arnoldkydim_chaotic} for larger values of $\tau_0$, we follow the argumentation on disordered circle maps in the limit $L\to\infty$ found in \cite{muller-bender_suppression_2022}.
There the finite-time Lyapunov exponent $\lambda_{K_L}[R_\omega] = \frac{1}{K_L} \sum_{k=0}^{K_L-1} \ln(1-\dot{\tau}_\omega(\theta_k))$ of an orbit $\mathcal{O}_L=\{ \theta_k \}_k=0,1,\dots,K_L-1$ that passes the phase space of the circle map once is considered.
Assuming that the minimal delay $\tau_{\text{min}} = \min_{\omega,t} \tau_\omega(t) $ is larger than the correlation length $l$ of the delay $\tau_\omega(t)$, $\tau_{\text{min}}>l$, one finds for the delays defined in Sec.~\ref{sec:delay_definition} that the increments $\tau_\omega(\theta_k)$ and $\tau_\omega(\theta_{k+1})$ are nearly independent.
More precisely, if the tails of the the Gaussian $G_\varsigma(t)$ in Eq.~\eqref{eq:random_delay} are neglected and therefore the Gaussian is replaced by a distribution with finite support $[-l,l]$ the increments are strictly independent.
In this case, the circle map with quenched disorder defined by Eq.~\eqref{eq:disordered_circle_map} is stochastically equivalent to the circle map with annealed disorder, where the disorder changes with each iteration $\omega=\omega_k$.
In the literature, such maps are known as ``random dynamical system on the circle'' \cite{malicet_random_2017,malicet_lyapunov_2021} (see \cite{muller-bender_suppression_2022} for more references) and it is known that their Lyapunov exponent is strictly negative if the realizations of the circle maps obtained from the disorder realizations $\omega_k$, almost surely, do not preserve the same measure \cite{malicet_random_2017,malicet_lyapunov_2021}, which is a very general assumption.
This means that $\lambda_{K_L}[R_\omega]$ converges to a unique, strictly negative value $\lambda[R]$ in the limit $L\to\infty$, which is associated with $K_L\to\infty$ as the length $K_L$ of the orbit $\mathcal{O}_L$ grows with the system size, $K_L \sim L/T$, where $T$ is the limit of the average roundtrip time, Eq.~\eqref{eq:avroundtriptime}, for $L\to\infty$.
For finite $L$, the authors of \cite{muller-bender_suppression_2022} further consider the distribution of the Lyapunov exponent of the disordered circle map, Eq.~\eqref{eq:disordered_circle_map}, with respect to the disorder.
Using the approximate independence of subsequent increments $\tau_\omega(\theta_k)$ and the central limit theorem, they find that the probability of generating a disorder realization with a Lyapunov exponent larger than $\lambda_{\text{cutoff}}$ with $\lambda_{\text{cutoff}} > \lambda[R]$ vanishes for increasing $L$ with
\begin{equation}
\label{eq:fracquasiperioasym}
	P(\lambda[R_\omega] > \lambda_{\text{cutoff}}) \sim (L/T)^{-1/2}\,e^{-c^2\,L/T} 
\end{equation}
where we have $c=(\lambda_{\text{cutoff}}-\lambda[R])/\sqrt{2\,\sigma_0^2}$ and $\sigma_0$ depends on the definition of $\tau_\omega(t)$.
Closing the gap between $L\to\infty$ and $L=\infty$, this further verifies that the Lyapunov exponent $\lambda[R]$ of the access map for random delay, which corresponds to $L=\infty$, is strictly negative for larger $\tau_0$ such that $\tau_{\text{min}}>l$.
It follows that we have $\lambda[R]<0$ and therefore dissipative delays are found in almost the whole delay parameter space which has consequences on the dynamics of the delay system that are described in the preceding section.
Sawtooth-shaped delays as considered in Fig.~\ref{fig:arnoldkydim}(b,c) and Fig.~\ref{fig:arnoldkydim_chaotic} lead to strongly negative exponents $\lambda[R]$, which is ideal for observing laminar chaos.
This is because in a large fraction of the phase space of the access map one has strongly negative values $\ln[\dot{R}(t)]=\ln[1-\dot{\tau}(t)] \approx \log(1-A) < 0$, whereas $\ln[\dot{R}(t)]$ is positive only at the transition between two sawtooths.
The transitions can be arbitrarily short or even of measure zero as in the case of the unsmoothed sawtooth delay, which is obtained in the limit $\varsigma \to 0$ in Eq.~\eqref{eq:random_delay} with Eq.~(\ref{eq:randomdelay_onoffsaw},\ref{eq:randomdelay_randdursaw}), where one has $\lambda[R]=\log(1-A)$.
The argumentation in this paragraph also holds for chaotically time-varying delays as long as the delay generating process only shows short-range correlations.
Questions remain for chaotic delays with long-range correlations such as partially predictable chaos \cite{wernecke_how_2017} and weak chaos \cite{grossmann_long_1985,bel_weak_2006,korabel_infinite_2012}, which is found for example in Eq.~\eqref{eq:chaotic_delay} at the intermittent transition to turbulence for values of $\rho$ close to $166$ \cite{manneville_intermittency_1979,pomeau_intermittent_1980}.

In summary, we have demonstrated that both random and chaotically time-varying delays can be classified by the dynamics of the access map, where the classification is unique in the sense that almost all realizations lead to the same values of the dynamical quantities if the delay parameters $\tau_0$ and $A$ are fixed.
In practice, the access map Lyapunov exponent $\lambda[R]$, Eq.~\eqref{eq:lyapunov_exponent}, can be used to classify the delay, where $\lambda[R]<0$ is characteristic for a dissipative delay and $\lambda[R]=0$ is characteristic for conservative delays, except limiting cases which are of measure zero in parameter space.
It follows that, in the same way known for (quasi-)periodic delays, chaotic dynamics of Eq.~\eqref{eq:sys} can be classified via the conditions for laminar and generalized laminar chaos, Eqs.~(\ref{eq:lamchaos_crit},\ref{eq:genlamchaos_crit}) in Sec.~\ref{sec:review}.
If the criterium for laminar chaos is fulfilled, the dynamics of the limit map given by Eq.~\eqref{eq:limit_map}, which holds for infinite $\Theta$, develops the nearly constant laminar phases that are characteristic for laminar chaos.
As a result of the assumed monotonicity of $R(t)=t-\tau(t)$, a laminar phase inside the solution segment $z_k(t)$ is mapped to a laminar phase in the solution segment $z_{k+1}(t)$, where the times at which a laminar phase is initiated and terminated are mapped from segment to segment by the (inverse) access map, Eq.~\eqref{eq:2d_map_r}.
While this leads to a (quasi-)periodic dynamics of the durations of the laminar phases for (quasi-)periodic delays, a random or chaotic dynamics is found for random and chaotically time-varying delays, respectively.
The levels of the laminar phases are connected by the one-dimensional map, Eq.~\eqref{eq:2d_map_f}, that is defined by the feedback nonlinearity of the delay system, Eq.~\eqref{eq:sys}.
For finite $\Theta$, where the smoothing kernel in the solution operator in Eq.~\eqref{eq:soluop} has a nonzero width $(1/\Theta)$, the laminar phases persist for $\Theta$ large enough such that the kernel width is much smaller than the time scale of the laminar phases developed by the limit map.
From this point of view, we found a straightforward generalization of the theory on (generalized) laminar chaos, where all of the features known from systems with (quasi-)periodic delay are present.
In stark contrast to the latter systems, low-dimensional (generalized) laminar chaos dominates almost the whole delay parameter space for random and chaotic delay variations (compare bottom panels of Figs.~\ref{fig:arnoldkydim}-\ref{fig:arnoldkydim_chaotic}), which follows from the strict negativity of the access map Lyapunov exponent, given that $A>0$ and $\tau_0$ large enough such that the minimal delay $\tau_{\text{min}}$ is larger than the correlation length of the delay variation.
Another major difference is found in the laminar chaotic time series itself, where random and chaotic delay variations lead to a time dependent number of laminar phases per solution segment $z_k(t)$, whereas this number is constant for (quasi-)periodic delays.
This feature requires an update of the time series analysis toolbox for the detection of laminar chaos, which is done in the next section.

\section{Features of laminar chaos and their detection}
\label{sec:test}

In the following we describe the characteristic features of laminar chaos in systems with random and chaotically time-varying delay.
We show that these features and therefore laminar chaos can be detected in time series without any knowledge of the system.
Therefore we generalize the time series analysis toolbox introduced in \cite{muller-bender_laminar_2020} and extended in \cite{muller-bender_pseudolaminar_2023}.
While we only use a time series generated by our system with chaotically time-varying delay for the numerical examples, the toolbox has been successfully tested for all of the delays considered above.
To demonstrate how our method can distinguish between laminar chaos and chaotic laminar dynamics generated by other mechanisms, we apply the method to a system that shows a generalization of the so-called \emph{pseudolaminar chaos} introduced in \cite{muller-bender_pseudolaminar_2023}.

\subsection{Test for laminar chaos}
\label{sec:test_laminar}

The main feature of laminar chaos is given by the nearly constant laminar phases, whose levels dynamics is governed the one-dimensional map given by Eq.~\eqref{eq:2d_map_f}.
Considering the simplest case, the dynamics of the levels of subsequent laminar phases is given by $z_{n+1}=f(z_n)$, where $z_n$ is the level of the $n$th laminar phase of a time series.
In this case, each solution segment $z_k(t)$ contains one laminar phase.
As shown in \cite{muller_laminar_2018,muller-bender_resonant_2019,muller-bender_laminar_2023}, the general behavior depends on the map given by Eq.~\eqref{eq:2d_map_r}, which is the inverse of the access map $t'=R(t)=t-\tau(t)$.
For periodic and quasiperiodic delays $\tau(t)$, one generally has a constant number $p$ of laminar phases per solution segment $z_k(t)$, such that the levels of the laminar phases of a time series fulfill $z_{n+p}=f(z_n)$.
Using these properties, the original test for laminar chaos basically consists of two steps \cite{muller-bender_laminar_2020,muller-bender_pseudolaminar_2023}.
Given a chaotic time series, the laminar phases are identified and their levels $z_n$ are extracted.
Finally, the return maps are generated by plotting the points $(z_n,z_{n+p})$ for different integers $p>0$, where $p$ is fixed for each return map.
If one finds values $p$, such that the points $(z_n,z_{n+p})$ resemble a one-dimensional function, the dynamics is classified as laminar chaos.
For the lowest value $p=p_{\text{true}}$, this function is simply the nonlinearity $f$ of the feedback of the delay system given by Eq.~\eqref{eq:sys} and $p_{\text{true}}$ is then the correct value for the number of laminar phases per solution segment $z_k(t)$.
If no $p$ is found such that the return map resembles a function, it can be concluded that a different mechanism generated the observed dynamics, since, as demonstrated in \cite{hart_laminar_2019,muller-bender_laminar_2020} laminar chaos can be identified by reconstructing the nonlinearity $f$ even under the influence of strong dynamical noise, where no laminar phases are apparently visible in the time series.
In the following, we demonstrate that the original test for laminar chaos fails for random and chaotically time-varying delay since the number of laminar phases per solution segment varies, $p=p_n$.
Therefore after extracting the laminar phases from the time series, in an additional step the sequence of the $p_n$ has to be found such that $(z_n,z_{n+p_n})$ resemble a one-dimensional function to classify the dynamics as laminar chaotic.
It turns out that the sequence $p_n$ can be extracted from the durations $\delta_n$ of the laminar phases using clustering methods.

\begin{figure}
    \centering
    \includegraphics[width=0.5\textwidth]{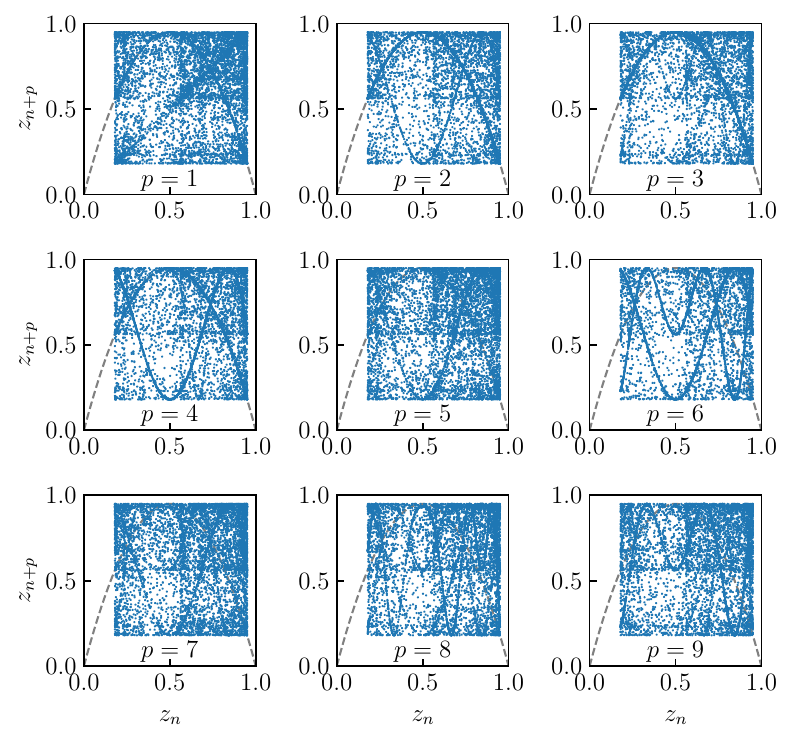}
    \caption{Return maps of the levels of the laminar phases.
        The level $z_{n+p}$ of the $(n+p)$th laminar phase is plotted over the level $z_n$ of the $n$th laminar phase for different $p$.
        The $z_n$ were extracted from a laminar chaotic time series of Eq.~\eqref{eq:sys} with the chaotically time varying delay shown in Fig.~\ref{fig:chaoticdelay}(b), where we set $\tau_0=3$ and $A=0.9$.
        While non of the return maps resembles the nonlinearity of Eq.~\eqref{eq:sys} (dashed lines), remnants of the nonlinearity are visible for $1 \leq p \leq 4$, which indicates that $p$ is time-dependent and varies between these values for the given parameters.
    }
    \label{fig:levels_chaoticdelay}
\end{figure}

\begin{figure}
    \centering
    \includegraphics[width=0.5\textwidth]{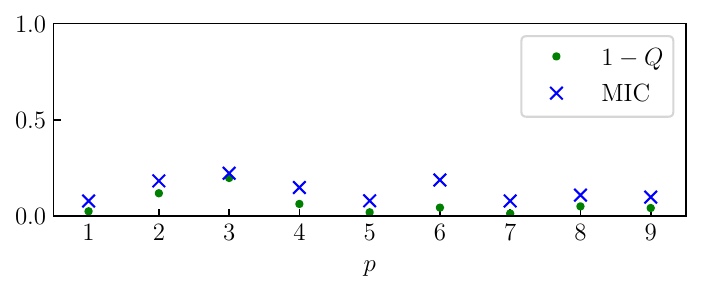}
    \caption{Analysis of the nonlinear correlations in the return maps in Fig.~\ref{fig:levels_chaoticdelay} of a laminar chaotic time series.
    A value of ${(1-Q)}$ and the MIC close to one would indicate strong nonlinear correlations as one would expect for at least one value of $p$ in the case of laminar chaos in systems with (quasi-)periodic delay.
    For random and chaotically time-varying delays, much smaller values are observed for all $p$, which indicates only weak correlations.
    This supports our conjecture that $p$ is time-dependent in this case.
    }
    \label{fig:lamchaostest_chaoticdelay}
\end{figure}

\begin{figure}
    \centering
    \includegraphics[width=0.5\textwidth]{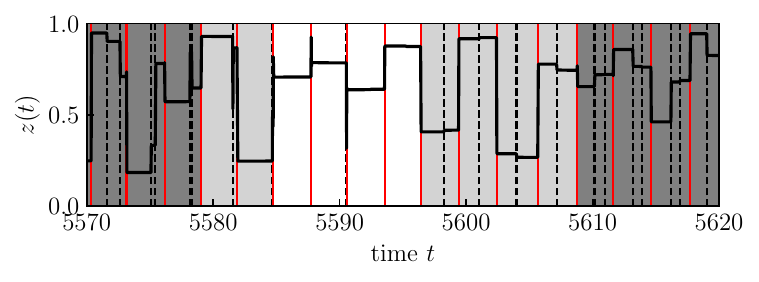}
    \caption{The number $p=p_k$ of laminar phases per state interval $(t_{k-1},t_k]$ varies from state interval to state interval.
        Laminar chaotic time series of Eq.~\eqref{eq:sys} with the chaotically time varying delay shown in Fig.~\ref{fig:chaoticdelay}, where we set $\tau_0=3$ and $A=0.9$.
        The boundaries of the laminar phases and of the state intervals are indicated by the dashed and red solid lines, respectively.
        In this snippet of the time series $p=p_k$ takes the values one, two, and three indicated by the background colors white, light gray, dark gray, respectively.
    }
    \label{fig:trajectory_chaoticdelay_p}
\end{figure}

\begin{figure}
    \centering
    \includegraphics[width=0.5\textwidth]{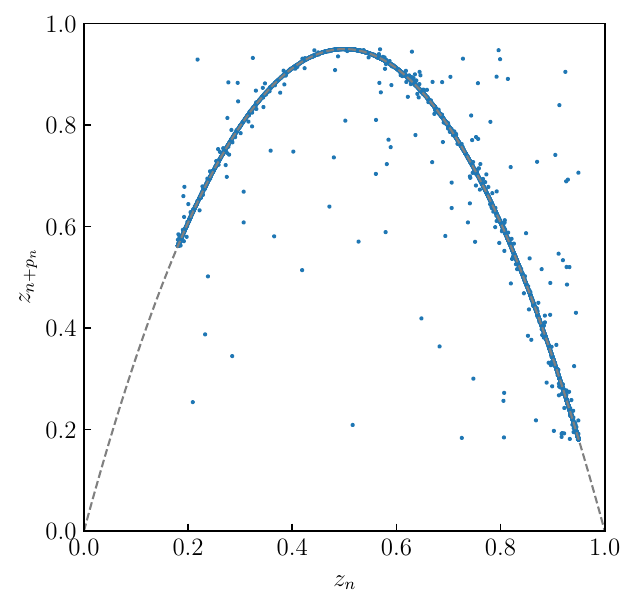}
    \caption{Return map of the levels $z_n$ of the laminar phases with varying $p=p_n$.
        The sequence of the $p_n$ was reconstructed using the knowledge of the time-varying delay (see text).
        The nonlinearity (dashed line) is well resembled, which is confirmed by the quantities $(1-Q) \approx 0.98$ and $\text{MIC} \approx 1.00$, which are close to one indicating strong nonlinear correlations as expected for laminar chaos.
    }
    \label{fig:retmap_chaoticdelay_truep}
\end{figure}

\begin{figure}
	\centering
	\includegraphics[width=0.5\textwidth]{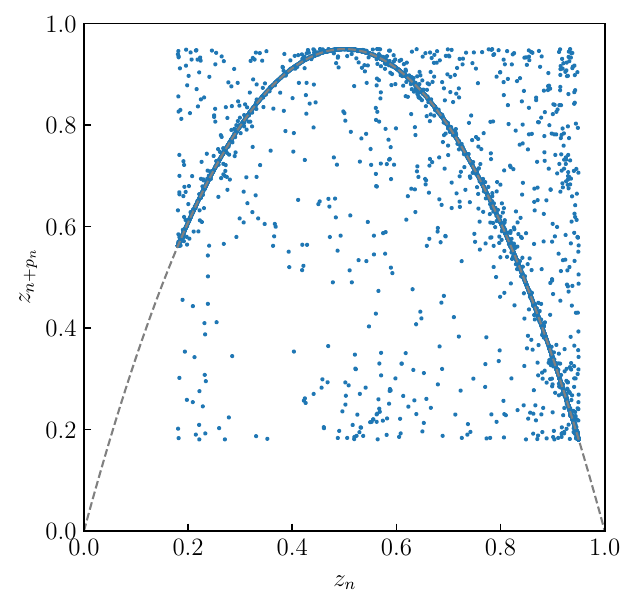}
	\caption{Return map of the levels $z_n$ of the laminar phases, where the sequence of the $p_n$ was extracted from the time series without knowing the system but using the clusters identified in Appendix~\ref{sec:appendix}, Fig.~\ref{fig:durations_pfromdata_chaoticdelay}.
		The nonlinearity (dashed line) is well resembled, but we find more outliers compared to Fig.~\ref{fig:retmap_chaoticdelay_truep}, where the delay function was used to determine the sequence of the $p_n$.
		This is also reflected by slightly lower values of the quantities $(1-Q) \approx 0.79$ and $\text{MIC} \approx 0.77$, which nevertheless indicate strong nonlinear correlations as expected for laminar chaos. 
	}
	\label{fig:retmap_chaoticdelay_fromdata}
\end{figure}

In the following we naively apply the original toolbox to a time series of a system with chaotically time-varying delay, where we choose the same parameters as in Fig.~\ref{fig:arnoldkydim_chaotic} and Fig.~\ref{fig:trajectories}(d) \footnote{The time series has a total length of $10^4$ time units, where we skip the first $10^3$ time units to let the transients relax.
}.
We first identify the laminar phases and extract their levels $z_n$
\footnote{To identify the laminar phases, we first detect the bursts between the laminar phases, where a burst is assumed when the absolute valued derivative $|\dot{z}(t)|$ of the time series exceeds the threshold $0.1$ and the duration between the end of the preceding burst and the current burst candidate is larger than $0.05$.
So a laminar phase is detected if $|\dot{z}(t)|$ is close to zero for a duration that is at least one magnitude larger than the characteristic time scale $1/\Theta$ with $\Theta=200$.
The burst position is then given by the onset of the burst, where $|\dot{z}(t)|$ starts to exceed the threshold.
The level $z_n$ of the $n$th laminar phase is defined as the value of $z(t)$ in the middle between two subsequent burst positions.}.
Then the return map of the laminar phases is plotted, which should reveal the nonlinearity $f$ of the delay system if laminar chaos is present.
However, this concept is only partially applicable for chaotically time-varying delay as shown in Fig.~\ref{fig:levels_chaoticdelay}.
The return maps clearly show remnants of the nonlinearity $f$ but with a noisy background and with remnants of higher iterates $f^l$.
While for periodic and quasiperiodic delay the first iteration of the map $z'=f(z)$ appears for a unique value $p=p_{\text{true}}$, it is found in Fig.~\ref{fig:levels_chaoticdelay} for $p=1,2,3,4$ (dashed line).
This indicates that the number $p_{\text{true}}$ of laminar phases per solution segment is in general not constant for random and chaotically time-varying delay.
As done in \cite{muller-bender_pseudolaminar_2023}, we now quantify the nonlinear correlations of these return maps, where strong nonlinear correlations are an indicator for the presence of laminar chaos generated from systems with periodic or quasiperiodic delay.
For that we compute the \emph{maximal information coefficient (MIC)} introduced in \cite{reshef_detecting_2011} using the Python package \emph{minepy} \cite{albanese_minerva_2013} and the quantity $(1-Q)$, which relates local fluctuations of the return map to global fluctuations of the levels $z_n$
\footnote{Following the idea of the CANOVA method \cite{wang_efficient_2015}, we relate local fluctuations of the return maps of the $z_n$ to their global fluctuations.
Therefore we sort the sequence of points $(z_n,z_{n+p})$ by the first component leading to the sequence of points $(v_m,w_m)$.
The quantity $Q$ is then defined by $Q = \frac{\text{Var}(w_{m+1}-w_m)}{2\, \text{Var}(w_m)}$, where $\text{Var}(w_m)$ is the sample variance of all values $w_m$.
More details can be found in \cite{muller-bender_pseudolaminar_2023}.
}.
The results for our system with chaotic delay are shown in Fig.~\ref{fig:lamchaostest_chaoticdelay}, where values close to $0$ or $1$ belong to weak or strong correlations, respectively.
Since both quantities are rather closer to $0$ than to $1$, the classical test for laminar chaos fails.
The reason for that is illustrated in Fig.~\ref{fig:trajectory_chaoticdelay_p}, where part of the analyzed time series is shown together with the numerically detected boundaries of the laminar phases and with the boundaries of the state intervals $\mathcal{I}_k = (t_{k-1},t_k] = (t_{k}-\tau(t_k),t_k]$, which are the domains of the solution segments $z_k(t)$ representing the memory of the delay system at time $t=t_k$.
Highlighted by the different shadings, the number $p=p_k$ of laminar phases per solution segment varies, so that the original test for laminar chaos fails as $p$ is assumed constant.
If the delay function $\tau(t)$ is known, it is straightforward to obtain the sequence $p_n$ such that the points $(z_n,z_{n+p_n})$ resemble the nonlinearity $f$.
Therefore we assign to the $n$th laminar phase a time $\tilde{t}_n$ equal to the numerically detected center of the phase.
Using the monotonicity of the access map, we know that the $n$th laminar phase is uniquely mapped to the $(n+p_n)$th laminar phase.
It follows that the correct $p_n$ with $z_{n+p_n}=f(z_n)$ is found if $R(\tilde{t}_{n+p_n})$ is inside the domain of the $n$th laminar phase.
Applying this approach to our exemplary time series gives us the associated sequence $p_n$.
However, for some $n$ no or no unique $p_n$ can be detected.
This happens when the plateau detection algorithm fails to detect a burst between two laminar phases, when spurious laminar phases are detected, or naturally when the number of laminar phases per solution segment changes, $p_{n+1} \neq p_n$ (see transitions between the regions with different color in Fig.~\ref{fig:trajectory_chaoticdelay_p}). 
Plotting the points $(z_n,z_{n+p_n})$ while omitting points with missing $p_n$ results in the return map shown in Fig.~\ref{fig:retmap_chaoticdelay_truep}, clearly reproducing the nonlinearity except for some outliers and showing strong nonlinear correlations as expected for laminar chaos.

\begin{figure}[h]
    \centering
    \includegraphics[width=0.47\textwidth]{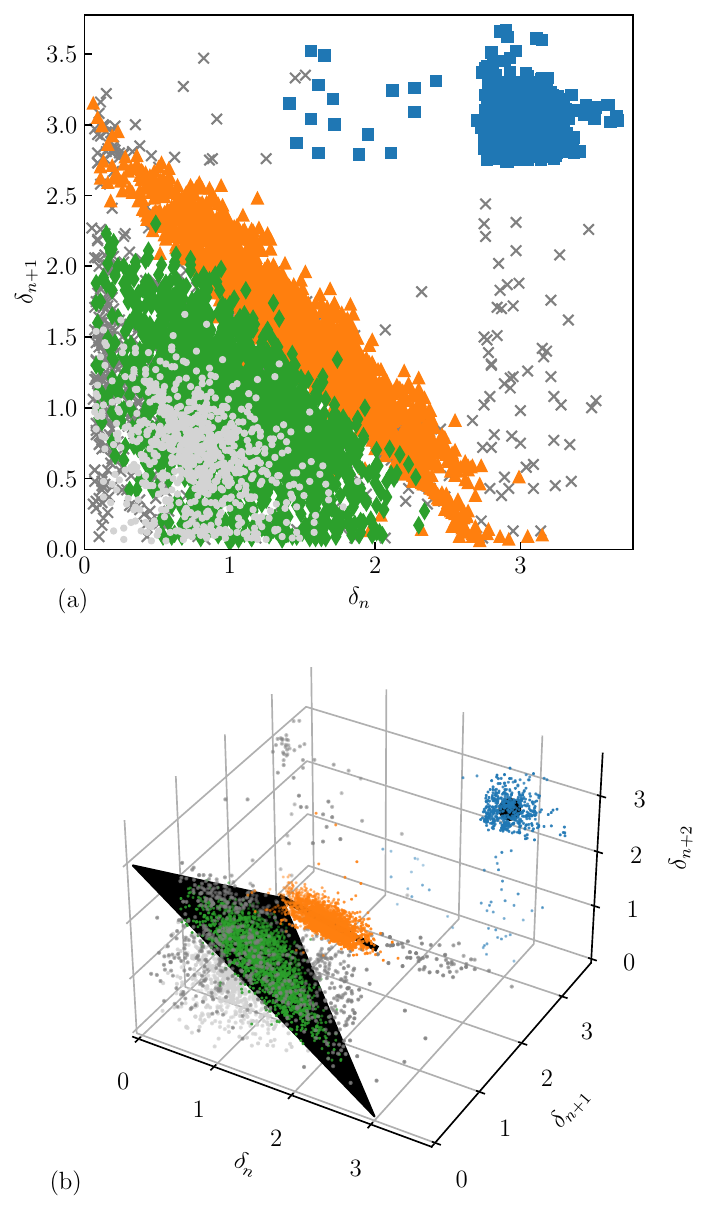}
    \caption{Mechanism behind laminar chaos leads to characteristic structures in $d$-dimensional return maps of the durations of the laminar phases with 
        (a) $d=2$, (b) $d=3$.
        Tuples of durations $(\delta_n,\delta_{n+1},\dots)$ of laminar phases with the same number $p_n=p_{n+1}=\dots=p$ of laminar phases inside the solution segments belong to the same point cluster.
        Depending on $p$ these clusters scatter around $(p-1)$-dimensional affine subspaces, which can be used to reconstruct the sequence $p_n$ from time series without knowing the system (see text).
        In (a), $p=1$, $p=2$, $p=3$, $p>3$ are represented by the blue squares, orange triangles, green diamonds, and light gray dots, respectively.
        If no unique $p$ is numerically found, dark grey crosses are used.
        In (b), the color coding is same and the $(p-1)$ subspaces defining the shape of the cluster are illustrated for $p=1,2,3$ by the black dot, line, and plane, respectively.
    }
    \label{fig:durations_chaoticdelay}
\end{figure}

If the system and therefore the time varying delay $\tau(t)$ is unknown, the sequence $p_n$ in principle can be reconstructed from the durations $\delta_n$ of the laminar phases using the $d$-dimensional return maps given by the points $(\delta_n,\delta_{n+1},\dots,\delta_{n+(d-1)})$.
In Fig.~\ref{fig:durations_chaoticdelay}, the two and three-dimensional return maps of the durations $\delta_n$ of the laminar phases are shown, which were computed by subtracting subsequent burst positions obtained from our exemplary time series.
The color coding of the values of the $p_n$ from our preceding analysis reveals characteristic structures that can be exploited for obtaining the $p_n$ even when we do not know anything about the system.
The structures arise from this trivial fact:
The sum of the durations of the laminar phases inside a solution segment are equal to the length of its domain, which is the state interval whose length equals the delay.
As the delay is bounded this leads to clustering around $(p-1)$-dimensional affine subspaces (see Appendix~\ref{sec:appendix} for details), which are illustrated in black in Fig.~\ref{fig:durations_chaoticdelay}(b).
Via the proof-of-concept clustering algorithm provided in Appendix~\ref{sec:appendix}, these clusters can be identified and the value of $p_n$ is assigned by looking to which cluster the point $(\delta_n,\delta_{n+1},\dots,\delta_{n+(d-1)})$ belongs.
Plotting $(z_n,z_{n+p_n})$ finally reveals the nonlinearity $f$ of the delay system as shown in Fig.~\ref{fig:retmap_chaoticdelay_fromdata}.
However, the nonlinear correlations are slightly lower than the values obtained when using the knowledge of the time-varying delay, compare Fig.~\ref{fig:retmap_chaoticdelay_truep}.
This is caused by the partial overlap of the point clusters for larger values of $p$, for instance, close to the point $(\delta_n,\delta_{n+1},\delta_{n+2}) = (0,T,0)$, where the subspace for $p=2$ (line) intersects the subspace for $p=3$ (plane).
Also the points in between the dense point clouds lead to wrong values $p_n$ since by the algorithm they are assigned to the clusters however they do not meet the assumption that $p$ must be equal for all laminar phases contributing to the point $(\delta_n,\delta_{n+1},\dots,\delta_{n+(d-1)})$ to be close to one of the affine subspaces.
This probably can be improved by updating the proof-of-concept algorithm used here, for instance, by using duration return maps with different dimensions $d$.
Then larger values of $p_n$ can be identified in higher-dimensions, where lower dimensions are more suitable for lower values of $p$ since $p_n$ has to be constant for at least $d$ laminar phases to lie close to one of the subspaces, which becomes less likely if $d$ is increased.

\subsection{Identification of pseudolaminar chaos}
\label{sec:test_plc}

In our previous work \cite{muller-bender_pseudolaminar_2023}, we considered a simple Lorenz-like system that constitutes an idealized model of one-dimensional motion of an active wave-particle entity -- a self-propelled droplet walking on a vibrating fluid bath. In this simple model, we observed pseudolaminar chaotic diffusion of the particle position in space when one of the internal parameters of the dynamical system was allowed to vary periodically, resulting in periodically driven on-off intermittency whose integrated signal gave the particle's position exhibiting pseudolaminar chaos. By adding a harmonic potential to the system, we have been able to keep the particle motion bounded and found trajectories that display features of pseudolaminar chaos with chaotically varying laminar phases when the internal parameter is allowed to fluctuate chaotically based on an independently driven chaotic system. We start by briefly reviewing the physical system and provide equations of motion.

A millimeter-sized drop of silicone oil can walk horizontally while bouncing vertically on a vertically vibrating bath of the same liquid~\cite{couder_bouncing_2005,valani_superwalking_2019}. The droplet on each bounce generates localized standing waves that decay slowly in time, and the droplet interacts with these self-generated waves on subsequent bounces to propel itself horizontally. We call such a walking droplet an \emph{active wave-particle entity} and it has the following three key features: (i) the droplet and the wave co-exist as a wave-particle entity; the repeated bouncing of the droplet sustains the damped waves which in turn guide the walking motion of the droplet, (ii) the system is active in the sense of an active particle~\cite{ramaswamy_mechanics_2010} since it absorbs energy from the vibrating bath and converts it into self-propulsion, and (iii) the system is non-Markovian and has path memory since the waves generated by the droplet decay very slowly in time, so the droplet is not only influenced from its most recent wave, but also from the waves generated in the distant past. By averaging over the fast time scale of periodic vertical bouncing, one obtains an integro-differential trajectory equation for horizontal walking motion~\cite{oza_trajectory_2013}. This trajectory equation for walking motion in $1$D in dimensionless form for a particle generating idealized cosine waves and confined to a harmonic potential takes the form of Lorenz-like nonlinear ordinary differential equations (ODEs) as follows~\cite{molacek_bouncing_2013,oza_trajectory_2013,durey_bifurcations_2020,valani_unsteady_2021,valani_lorenz-like_2022}:
\begin{align}\label{eq: WPE harmonic}
    \dot{x}&=X\\ \nonumber
    \dot{X}&=\sigma_0 (Y-X-k x)\\ \nonumber
    \dot{Y}&=-Y+X (r-Z)\\ \nonumber
    \dot{Z}&=-Z+XY \nonumber
\end{align}
Here $x$ is the particle position, $X$ is the particle velocity and dots denote derivatives with respect to time. The variables $Y$ and $Z$ are related to the wave-memory force on the particle from its history of self-generated waves~(see \citet{valani_lorenz-like_2022}). The parameter $\sigma_0^{-1}$ represented a dimensionless particle mass and the parameter $r$ represents a dimensionless wave-memory force coefficient. The constant $k$ is the spring constant forming an external harmonic potential. In summary, Eq.~\eqref{eq: WPE harmonic} describes the motion of a damped particle in a harmonic potential that generates slowly decaying cosine waves at each instant of time and propels itself based on the gradient of these self-generated waves. 

\begin{figure}
	\includegraphics[width=0.5\textwidth]{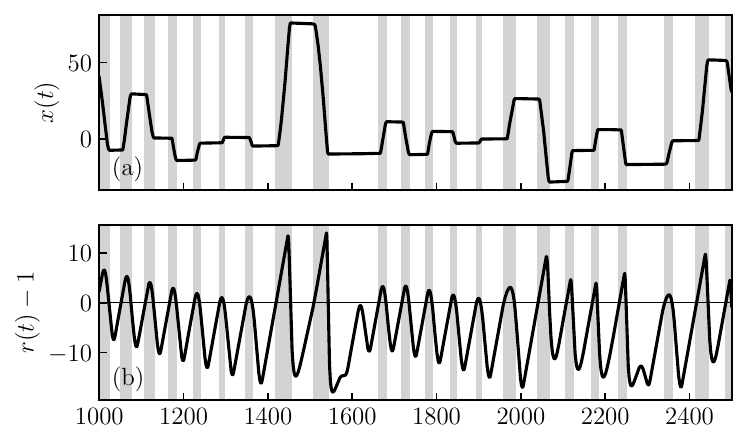}
	\caption{Mechanism behind pseudo-laminar chaos for the system in Eq.~\eqref{eq: WPE harmonic}.
    A laminar phase persists as long as the chaotically varying parameter approximately fulfills $r(t)<1$, corresponding to a stationary state of the dynamical system, and a deviation from the laminar phases is triggered when we have $r(t)>1$, corresponding to the instability of the stationary state.
    (a) Time series of the variable $x$ in the system of Lorenz-like ODEs in Eq.~\eqref{eq: WPE harmonic} with $r=r(t)$.
    (b) Shifted chaotically varying parameter $r(t)-1$. See also Supplemental Video 1.
	}
	\label{fig:trajectory_plc}
\end{figure}

We start by finding equilibrium states of the Lorenz-like system in Eq.~\eqref{eq: WPE harmonic}. We find that the only equilibrium state of the system is a stationary particle at the minima of the external harmonic potential given by $(x,X,Y,Z)=(0,0,0,0)$. By doing a linear stability analysis, we find that this state becomes unstable (undergoing a Hopf bifurcation) for $r>1+k\frac{\sigma_0}{1+\sigma_0}$. For larger values of $r$, complex set of global bifurcations take place, similar to the standard Lorenz system~\citep{lorenz_deterministic_1963}, and one gets chaotic dynamics.

We now consider the case where we allow the parameter $r=r(t)$ to vary chaotically.
In particular, we choose $r(t)=r_0+a\,h_4(\nu\,t)$.
The temporal variation $h_4$ is the last component of the chaotic system defined by Eq.~\eqref{eq:lorenz} in Sec.~\ref{sec:delay_definition}, where we use the parameters given there; the additional parameter $\nu$ enables us to set the time scale of the driving.
The dynamics of the system can be understood in terms of the variations in the value of $r(t)$. Since we fix $k=0.001$, approximately, for $r(t)<1$, the system will be in a stationary state and hence we would get laminar phases in the time series of particle position $x$, whereas for $r(t)>1$, the stationary state becomes unstable and the particle position undergoes in general chaotic motion within the harmonic potential at larger values of $r$, resulting in an irregular burst. At a later time, when $r(t)<1$, a new laminar phase will be created at this location. Hence, the start and the end of the laminar phases are correlated with the criteria $r(t)=1$.
This mechanism is illustrated in Fig.~\ref{fig:trajectory_plc}, where the parameters were set to $\sigma_0=5$, $k=0.001$, $r_0=-20$, $a=40$, and $\nu=0.012$, which will also be used in the following analysis.

\begin{figure}
    \centering
    \includegraphics[width=0.5\textwidth]{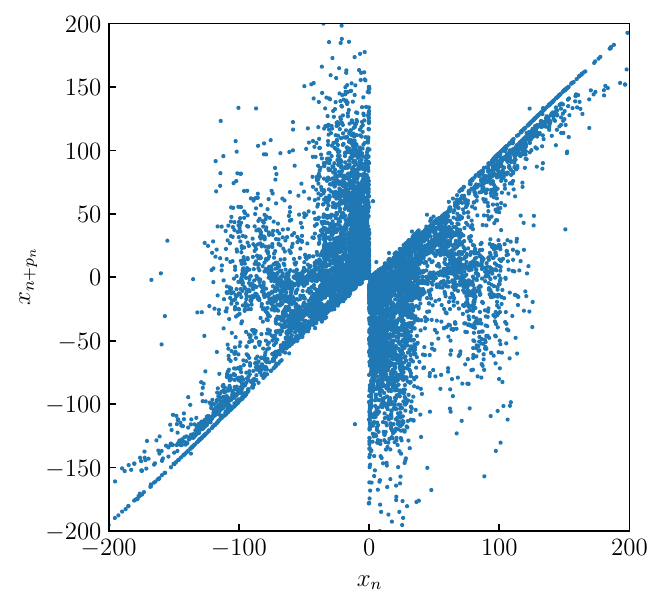}
    \caption{Return map of the levels $x_n$ of the laminar phases of pseudolaminar chaos, where the sequence of the $p_n$ was extracted using the clusters identified in Fig.~\ref{fig:durations_pfromdata_plc_chaoticparameter}.
    While the quantities $(1-Q) \approx 0.52$ and $\text{MIC} \approx 0.44$ indicate some nonlinear correlations, the points do not resemble a function, which is contradictory to laminar chaos.
    }
    \label{fig:retmap_plc_chaoticparameter_fromdata}
\end{figure}

We now apply our test for laminar chaos introduced in the previous section to an exemplary time series with the length of $10^6$ time units
\footnote{The time series was generated via the MATLAB function \texttt{ode45} while setting the options \texttt{'RelTol'} and \texttt{'AbsTol'} both to $10^{-10}$.}, where the first $1000$ time units are dropped to let the transients relax.
Further technical details can be found in Appendix~\ref{sec:appendix}.
We find that the quantities $(1-Q)$ and MIC take values around $0.5$ and therefore indicate some sort of correlations, which are caused by the harmonic potential pulling the particle position towards $x=0$.
However, the return map as shown in Fig.~\ref{fig:retmap_plc_chaoticparameter_fromdata} is far from resembling a one-dimensional map, which is contradictory to laminar chaos and is expected since this type of dynamics is generated by a completely different mechanism.
So the updated test for laminar chaos enables us to distinguish this type of pseudolaminar chaotic dynamics from laminar chaos if we do not only rely on the values of quantities like $(1-Q)$ and MIC but also take into account the actual structure of the return map of the levels.

We additionally performed numerical experiments with periodically varying $r(t)$, where the parameters were tuned such that the return map of the levels becomes as close as possible to a one-dimensional map and shows very strong nonlinear correlations.
It turned out that these correlations are very sensitive and vanish if one adds dynamical noise, with a very small noise strength, as we have found previously for pseudo-laminar chaotic diffusion generated by periodically driven on-off intermittency \cite{muller-bender_pseudolaminar_2023}. Since noise is always present in real world systems, it is unlikely that one finds an experimental time series showing pseudolaminar chaos that passes the test for laminar chaos.

\section{Summary}

We have generalized the theory of laminar chaos to systems with random and chaotically time-varying delay.
Surprisingly, for short-time correlated delays, low-dimensional (generalized) laminar chaos is observed in almost the whole delay parameter space spanned by the mean delay and the delay amplitude, whereas high-dimensional turbulent chaos is only observed if the minimum of the delay is of the order of the correlation length of the temporal delay variation or smaller.
Therefore, random and chaotically varying delays are in stark contrast to constant delays where only turbulent chaos is observed. The introduction of randomness or chaos in delay variation typically leads to a drastic reduction of the dimension of the chaotic attractor of the considered systems.

In addition to the random or chaotic variations of the duration of the nearly constant laminar phases, we found that the number $p_n$ of laminar phases inside the memory of the system varies so that the level $z_n$ of the $n$th laminar phase determines the level $z_{n+p_n}$ of the $(n+p_n)$th laminar phase by $z_{n+p_n}=f(z_n)$, where $f$ is the feedback nonlinearity of the system.
As the reconstruction of $f$ is crucial for the test of experimental time series for laminar chaos and the sequence of $p_n$ is unknown in this case, we introduced a proof-of-concept algorithm for the extraction of the sequence $p_n$ from the time series.
It relies on the identification of clusters in return maps of the durations of the laminar phases -- an additional characteristic feature of laminar chaos in systems with random and chaotically time-varying delay.
A comparison of the results of our algorithm applied to a laminar chaotic time series and a pseudolaminar chaotic time series generated by a by a Lorenz-like ODE system with chaotically varying parameters, indicated the strengths and the limitations of the algorithm.
While laminar chaos can be identified by a careful analysis of the output of the algorithm, it is not a zero-one test for laminar chaos, where one can rely on the value of a single scalar quantity.
If only the output quantification of the nonlinear correlations between the levels of the laminar phases are considered, errors in the cluster identification may lead to false negatives or correlations in pseudolaminar dynamics may lead to false positives. However, the latter is unlikely for experimental time series as pseudolaminar chaos is sensitive to noise such that correlations between the nearly constant phases are destroyed even for small noise strengths~\cite{muller-bender_pseudolaminar_2023}.

For random and chaotically time-varying delays with long-range correlations, we expect that the overall delay parameter space shows fractal structures since they form an intermediate case between regular delay variations and the short-time correlated delays considered here.
These will be investigated in a future publication in the context of delays that are both time and state-dependent, $\tau=\tau(t,z(t))$.
Long-range correlations can occur for instance in the case, where the time-dependence is periodic, $\tau(t+t_{\text{period}},z(t))=\tau(t,z(t))$, but the delay actually varies chaotically due to the chaotic variation of the state $z(t)$. 
Questions remain for weakly chaotic delays, which means that the delay generating process shows weak ergodicity breaking \cite{bel_weak_2006} such that infinite ergodic theory plays a role and therefore time averages are random variables \cite{korabel_infinite_2012}).
It would be interesting and nontrivial to investigate whether laminar chaos can exist in systems with such delays or the dynamics randomly switches between laminar chaos and turbulent chaos as observed in \cite{albers_weak_2024} for a weakly chaotic feedback nonlinearity or something completely unexpected happens.

Reviewing the results on periodic delay \cite{muller_laminar_2018}, quasiperiodic delay \cite{muller-bender_laminar_2023} and the results on random and chaotically time varying delay presented here, we see that the theory behind laminar chaos, its features and methods for their detection become more and more complicated as the complexity of the delay increases.
Further complexity is expected, when we consider more general systems than Eq.~\eqref{eq:sys}.
This basically sets the long-time goal of our research: to find a system independent, geometrical description of laminar chaos.
That should be possible as the main concept behind it, having a so-called dissipative delay, is independent of the specific system \cite{otto_universal_2017}.

\begin{acknowledgments}
D.MB. gratefully acknowledges funding by the Deutsche Forschungsgemeinschaft (DFG, German Research Foundation) - 456546951. R.V. acknowledges the support of the Leverhulme Trust [Grant No. LIP-2020-014]. 
We would also like to thank Tony Albers for thoroughly reading the manuscript and for his valuable suggestions.
All authors gratefully acknowledge the contributions of the late Günter Radons to the results presented here.
He initiated and supervised D.MBs. work on systems with time-varying delay and managed the DFG project on generalizations of laminar chaos until he passed away on July 20, 2024.
We miss him.
\end{acknowledgments}

\appendix
\section{Clusters of the durations of laminar phases and their detection}
\label{sec:appendix}

\begin{figure}[b]
	\includegraphics[width=0.5\textwidth]{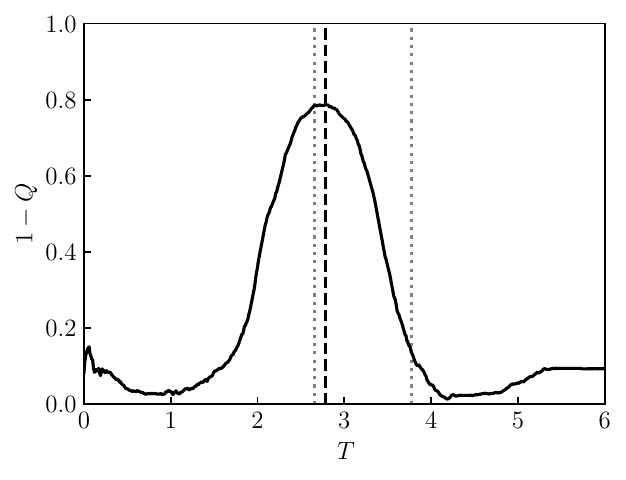}
	\caption{Quantity $(1-Q)$ as a function of the parameter $T$ of the clustering algorithm.
		For each $T$, the sequence $p_n$ is extracted from the $(d=4)$-dimensional durations return map.
		Then $(1-Q)$ is computed for the points $(x_n,x_{n+p_n})$.
		The maximum at $T=2.78$ (dashed line) is located within the range of the time-varying delay (dotted lines). 
	}
	\label{fig:score}
\end{figure}

\begin{figure}[b]
	\centering
	\includegraphics[width=0.5\textwidth]{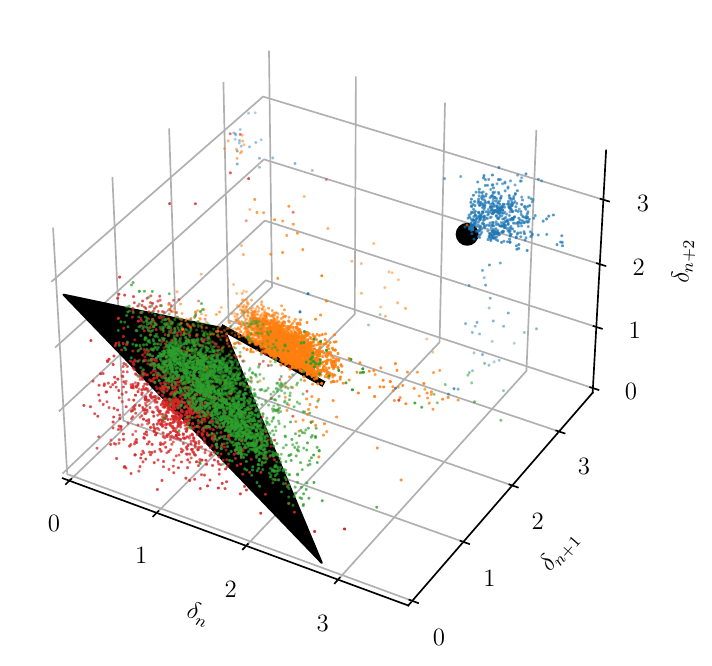}
	\caption{Numerically identified clusters from a $(d=4)$-dimensional return map of the durations of the laminar phases without knowing the system (see text), where $p=1$, $p=2$, $p=3$, $p=4$ are represented blue, orange, green and red, respectively.
		The estimated subspaces defining the shape of the cluster for $p=1,2,3$ are represented by the black dot, line and plane.
	}
	\label{fig:durations_pfromdata_chaoticdelay}
\end{figure}

\begin{figure}[b]
	\includegraphics[width=0.5\textwidth]{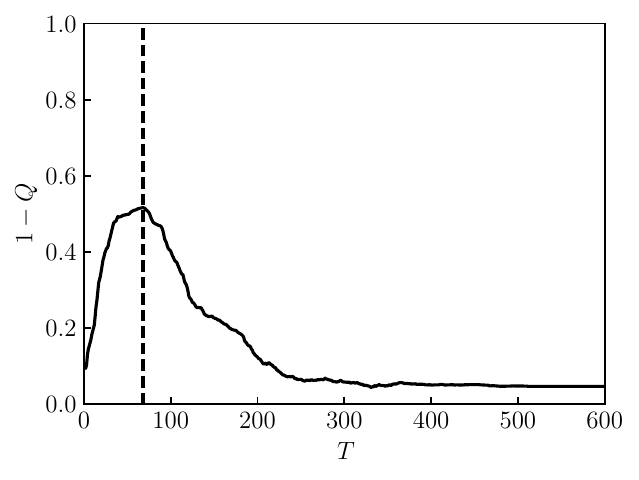}
	\caption{Quantity $(1-Q)$ as a function of the parameter $T$ of the clustering algorithm for pseudolaminar chaos, where the $(d=4)$-dimensional return map of the durations of the (pseudo)laminar phases was analyzed.
		The return maps of the levels and durations of the laminar phases for the parameter $T \approx 68$ (dashed line) that maximizes $(1-Q)$ are used to classify the type of chaotic dynamics.  
	}
	\label{fig:score_plc}
\end{figure}

\begin{figure}[b]
	\centering
	\includegraphics[width=0.5\textwidth]{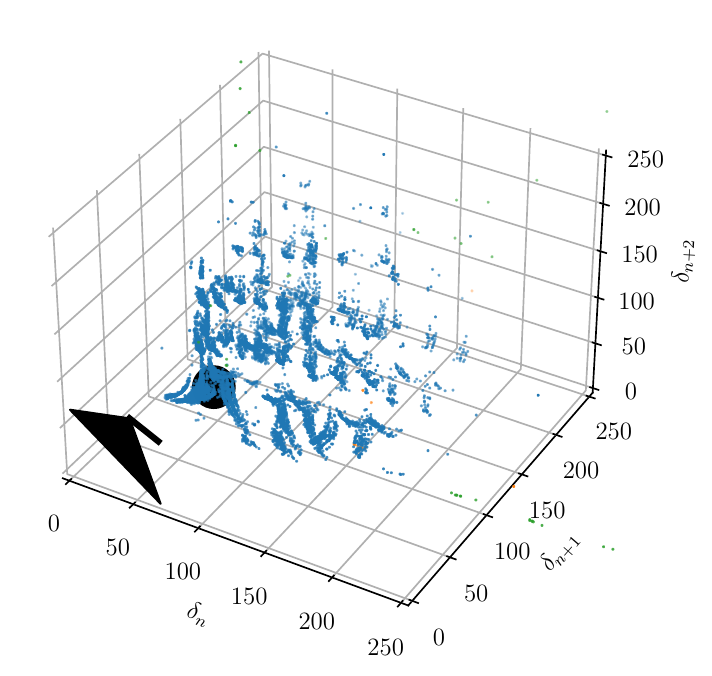}
	\caption{Numerically identified clusters from a $(d=4)$-dimensional return map of the durations of the laminar phases for pseudolaminar chaos.
		The vast majority of the points are assigned to the cluster with $p=1$ (blue), where only a few outliers were assigned to $p=2,3,4$ (orange, green, red).
		The estimated subspaces defining the shape of the cluster for $p=1,2,3$ are represented by the black dot, line and plane.
	}
	\label{fig:durations_pfromdata_plc_chaoticparameter}
\end{figure}

In the following we provide the technical details of the generalization of the toolbox for the detection of laminar chaos to system with random and chaotically time-varying delay.
A distinctive feature of laminar chaos is given by the fact that the levels $z_n$ are governed by the dynamics of the one-dimensional iterated map $z_{n+p}=f(z_n)$, where $p$ is the number $p$ of laminar phases per state interval.
Therefore the reconstruction of this map from a time series of an unknown system is crucial for the detection of laminar chaos.
While $p$ is constant for periodic and quasiperiodic delays and therefore easy to find, for random and chaotically time-varying delays it is time dependent, $p=p_n$, rendering trial-and-error approaches impossible. 
In the first part we present a systematic method how the number $p_n$ of laminar phases per state interval can be extracted from return maps of the durations $\delta_n$ of the laminar phases of a time series.
This method is benchmarked in the second part using time series that show laminar chaos and pseudolaminar chaos, where the assigned sequences of $p_n$ lead to return maps $z_{n+p}=f(z_n)$ that allow a clear distinction of these type of dynamics without knowing the system as demonstrated in Sec.~\ref{sec:test}

For the reconstruction of the number $p_n$ of laminar phases per state interval, we extract the sequence of durations $\delta_n$ of laminar phases from a time series, where $n$ is the number of the laminar phase in the time series.
We then exploit the geometrical structures that appear in $d$-dimensional return maps constructed from points $(\delta_n,\delta_{n+1},\dots,\delta_{n+(d-1)})$.
The structures arise from this trivial fact:
Given that a solution segment $z_k(t)$ contains only complete laminar phases, then the sum of the durations of these laminar phases are equal to the length of its domain, which is the state interval $\mathcal{I}_k=(t_k-\tau(t_k),t_k]$ and is of one delay length $\tau(t_k)$.
If there are $p$ laminar phases inside the solution segments, it follows that the components of each point $(\delta_n,\delta_{n+1},\dots,\delta_{n+(d-1)})$ of our $d$-dimensional return map fulfill $(d-p+1)$ equations given by
\begin{equation}
	\label{eq:durations_affsub}
	T_l=\sum_{m=l}^{l+(p-1)} \delta_{n+m}, \quad l=0,1,\dots,d-p.
\end{equation}
So each $T_l$ is simply the sum of $p$ subsequent durations starting with $\delta_{n+l}$.
Considering $T_l$ as constants and the durations $(\delta_n,\delta_{n+1},\dots,\delta_{n+(d-1)})$ as variables, Eqs.~\eqref{eq:durations_affsub} define a $(p-1)$-dimensional affine subspace of $\mathbb{R}^d$, where the $T_l$ define its offset from the origin.
Since we assumed that $p$ is equal for all laminar phases contributing to the point $(\delta_n,\delta_{n+1},\dots,\delta_{n+(d-1)})$, each $T_l$ is a length of a state interval and therefore is equal to a value of the time-varying delay $\tau(t)$.
This means that each vector composed of $d$ subsequent durations belongs to an $(p-1)$-dimensional affine subspace, where the offset of the subspace fluctuates according to the time-varying delay $\tau(t)$.
Averaging these fluctuations by replacing $T_l$ with the average roundtrip time of the feedback loop $T=T_\omega$ from Eq.~\eqref{eq:avroundtriptime} with $L\to\infty$ leads to the subspaces represented by the black dot, line, and plane in Fig.~\ref{fig:durations_chaoticdelay}(b), where the clusters with the corresponding values of $p=1,2,3$ scatter around.
For our chaotically time-varying delay with the parameters used here, we obtained an average roundtrip time of $T=T_\omega=3.0.$
The loose points away from these clusters belong to the case where $p$ is not equal for all laminar phases contributing to the point $(\delta_n,\delta_{n+1},\dots,\delta_{n+(d-1)})$, which happens at a transition from one value of $p_n$ to another, see Fig.~\ref{fig:trajectory_chaoticdelay_p}.

These structures are general in the sense that they are visible as long as the fluctuations of the delay $\tau(t)$ and therefore the fluctuations of the lengths of the state intervals are not too large such that the average roundtrip time $T$ and therefore the offset of the subspaces is finite, as well as the fluctuations around the subspace are small enough such that the clusters for different values of $p$ are distinguishable.
Therefore they can be used to extract the sequence $p_n$ from the time series by dividing the $d$-dimensional return map of the durations into such point clusters. 
After that, the value $p_n$ of the $n$th laminar phase can be determined simply by looking to which cluster the point $(\delta_n,\delta_{n+1},\dots,\delta_{n+(d-1)})$ belongs.

One possible method for solving this classification problem can be derived from the \emph{k-subspace clustering} method introduced in \cite{wang_k-subspace_2009}.
The points of the $d$-dimensional return map are classified by the Euclidean distance to the subspaces defined by Eq.~\eqref{eq:durations_affsub} with $T_l=T$.
In detail, $p_n$ is set of the value $p$ such that the distance between the corresponding subspace and the point $(\delta_n,\delta_{n+1},\dots,\delta_{n+(d-1)})$ is minimal.
Since $T$ is unknown, we scan for the value of $T$ that maximizes the nonlinear correlations of the return map given by the points $(z_n,z_{n+p_n})$ quantified by $(1-Q)$.

Such a scan computed from a time series of our exemplary laminar chaotic system considered in Sec.~\ref{sec:test_laminar} is shown in Fig.~\ref{fig:score}, where the algorithm was applied to a $4$-dimensional return map of durations such that $4$ clusters corresponding to $p=1,2,3,4$ were identified.
The cluster classification obtained at the value $T$ that maximizes $(1-Q)$ is shown in Fig.~\ref{fig:durations_pfromdata_chaoticdelay}.
It nicely coincides with the result shown in Fig.~\ref{fig:durations_chaoticdelay}(b), which was obtained using the knowledge of the time-varying delay, except that the estimate $T \approx 2.78$ differs from the expected value $T=3.0$ leading to the visible shift of the clusters relative to the subspaces.

Now we apply the exact same method to the pseudolaminar chaotic time series $x(t)$ analyzed in Sec.~\ref{sec:test_plc}.
We identify the laminar phases and extract the durations $\delta_n$ and the levels $x_n$ of the nearly constant phases.
Using the $(d=4)$-dimensional return map of the durations $\delta_n$, we try to identify the sequence of the $p_n$ by identifying the clusters that are associated to the specific values.
Therefore we vary the parameter $T$ that determines the location of the subspaces defining the shape of the clusters, compute the sequence $p_n$ from the results and compute the value of the quantity $(1-Q)$ for the resulting return map given by the points $(x_n,x_{n+p_n})$.
As shown in Fig.~\ref{fig:score_plc}, the nonlinear correlations quantified by $(1-Q)$ are maximized for $T=68$, where a step size of $\Delta T=1$ was used.
The three-dimensional return map of the durations together with the subspaces for $p=1,2,3$ is shown in Fig.~\ref{fig:durations_pfromdata_plc_chaoticparameter}.
The majority of points were assigned to the point cluster with $p=1$, where the remaining ones make of $0.6\%$ and belong to exceptional long nearly constant phases mostly exceeding the plot range of Fig.~\ref{fig:durations_pfromdata_plc_chaoticparameter}.
This means that the return map of the levels $x_n$ given by the points $(x_n,x_{n+p_n})$ is dominated by the relation between two subsequent nearly constant phases.

\bibliography{references}

\end{document}